\documentclass[acmsmall,authorversion]{acmart}

\AtBeginDocument{%
  \providecommand\BibTeX{{%
    \normalfont B\kern-0.5em{\scshape i\kern-0.25em b}\kern-0.8em\TeX}}}

\acmJournal{TOSEM}
\acmDOI{10.1145/3511805}

\usepackage{multirow, pifont, threeparttable, soul,xspace, multicol, lscape}
\newcommand{\ie}{i.e.\xspace}
\newcommand{\eg}{e.g.\xspace}
\def\acronym{\textsc{TePIA}\xspace}

\begin{document}


%\title[Information Attributes for Test Case Prioritisation]{Information Attributes for Test Case Prioritisation:
%Taxonomy, Applicability, and Machine Learning}
\title{A Taxonomy of Information Attributes for Test Case Prioritisation: Applicability, Machine Learning}

%\email{webmaster@marysville-ohio.com}
%\affiliation{%
%  \institution{Institute for Clarity in Documentation}
%  \streetaddress{P.O. Box 1212}

\author{Aurora Ram\'irez}
\authornote{Work partially done as visiting researcher at Chalmers University of Technology.}
\affiliation{%
  \institution{Departamento de Inform\'atica y An\'alisis Num\'erico, Universidad de C\'ordoba. Instituto Andaluz Interuniversitario de Data Science and Computational Intelligence (DaSCI)}
  \streetaddress{Campus de Rabanales, Edificio Marie Curie}
  \city{C\'ordoba}
  \country{Spain}
  \postcode{E-14071}
  }
\email{aramirez@uco.es}

\author{Robert Feldt}
\affiliation{%
  \institution{Software Engineering Division, Department of Computer Science and Engineering, Chalmers University of Technology}
  \streetaddress{Jupiter Building, Lindholmen Campus}
  \city{Gothenburg}
  \country{Sweden}
  \postcode{SE-412 96}
  }
\email{robert.feldt@chalmers.se}

\author{Jos\'e Ra\'ul Romero}
\affiliation{%
  \institution{Departamento de Inform\'atica y An\'alisis Num\'erico, Universidad de C\'ordoba. Instituto Andaluz Interuniversitario de Data Science and Computational Intelligence (DaSCI)}
  \streetaddress{Campus de Rabanales, Edificio Marie Curie}
  \city{C\'ordoba}
  \country{Spain}
  \postcode{E-14071}
  }
\email{jrromero@uco.es}

\renewcommand{\shortauthors}{Ram\'irez et al.}

\begin{abstract}
Most software companies have extensive test suites and re-run parts of them continuously to ensure recent changes have no adverse effects. Since test suites are costly to execute, industry needs methods for test case prioritisation (TCP). Recently, TCP methods use machine learning (ML) to exploit the information known about the system under test (SUT) and its test cases. However, the value added by ML-based TCP methods should be critically assessed with respect to the cost of collecting the information. This paper analyses two decades of TCP research, and presents a taxonomy of 91 information attributes that have been used. The attributes are classified with respect to their information sources and the characteristics of their extraction process. Based on this taxonomy, TCP methods validated with industrial data and those applying ML are analysed in terms of information availability, attribute combination and definition of data features suitable for ML. Relying on a high number of information attributes, assuming easy access to SUT code and simplified testing environments are identified as factors that might hamper industrial applicability of ML-based TCP. The \acronym taxonomy provides a reference framework to unify terminology and evaluate alternatives considering the cost-benefit of the information attributes.
\end{abstract}




\maketitle

%--------------------------------
% 1 - INTRODUCTION
%--------------------------------
\section{Introduction}\label{sec:intro}

Software testing is an essential activity in the software life cycle to ensure that the system meets the specification and is as free of errors as possible. As the software evolves, regression testing (RT) activities are planned to guarantee that changes in the system do not alter the behaviour of stable parts~\cite{Yoo12}. In modern scenarios of continuous integration and delivery, re-running the whole test suite for large-scale industrial systems is impractical~\cite{wikstrand2009dynamic,Minhas20}. Therefore, selecting and prioritising the most relevant test cases from extensive test suites becomes a core step of RT~\cite{PradoLima20}. Test case prioritisation (TCP) methods ---which often involves test case selection--- analyse information from the test cases and about their execution to decide which ones should be executed first. Thus, TCP aims at improving the global fault detection rate and reducing the time to detection, among other desired effects~\cite{Khatibsyarbini18}. Therefore, TCP requires organisations to keep information about their projects, testing practices, and testing artefacts, a process that may be difficult to reliably maintain over time. Necessarily, the effort needed to collect and maintain the information for TCP should also be balanced with the expected benefit in reduced costs associated to error detection and correction~\cite{Westland02}. This requires careful analysis of the industrial context, in particular since many TCP methods assume availability of information that is not relevant or measurable in practice~\cite{Ali19EMSE}.

A recent trend in TCP is the application of machine learning (ML)~\cite{CastroCabrera20,Pan21}, which has proven to be a suitable alternative to facilitate the automation of different testing tasks~\cite{Durelli19}. The strength of ML lies in its ability to discover patterns from data and make predictions, so the amount and quality of the input data are critical factors for success~\cite{Kotsiantis06, Briand08}. Lack of data is a recurrent problem when it comes to applying ML in software engineering~\cite{Menzies01}, including software testing~\cite{Menzies10,Ding20}. In the context of TCP, data quality is intrinsically linked to the identification of relevant information sources. Data acquisition and preparation become even more relevant when ML-based TCP methods are conceived for industrial application, as discussed in some recent case studies~\cite{Bus16,Ala18,Rop19}. In particular, the heterogeneity and scale of the systems, the presence of networking functionalities and hardware-dependent test cases, or the need of parsing execution traces illustrate this point. However, research on ML-based TCP is growing fast, boosted by easy access to repository information and increasingly complex ML algorithms and implementations\slash libraries able to deal with many attributes. Building predictive models without bearing the constraints and limitations of the testing environment in mind will render the results less useful and acceptable. The design of ML models applicable at industrial scale should thus be guided by the same principles as any other TCP method, and must take the industrial context, such as the availability of information, the costs of its acquisition as well as its maintenance, into account. Methods that require information that is costly to collect and\slash or maintain the quality of are not likely to be useful, regardless of its performance in an optimal situation.

The artefacts, \eg test cases, and processes, \eg test execution, from which information for TCP is derived are known as information sources, and have been analysed as part of literature reviews~\cite{PradoLima20} and discussions of industrially-relevant research~\cite{Ali19EMSE}. Information sources can provide diverse inputs for TCP methods depending on what is exactly measured in them. We will refer to these ``inputs'' as \emph{information attributes} throughout the paper, aiming at providing an in-depth evaluation of how the information is captured, transformed and combined in the context of automated TCP. For ML studies, we will use the commonly-accepted term \emph{data features} to refer to the inputs of the learning algorithm. This will allow us to study how information attributes are mapped into data features. Information attributes have already been classified in the aforementioned secondary studies from the perspective of continuous integration environments only~\cite{PradoLima20} or as part of the analysis of the RT industrial solution~\cite{Ali19EMSE}. Here, we focus on a more detailed analysis of their characteristics, presented in the form of a taxonomy, which allows us to discuss the implications of building data features for applicable ML-based TCP.   

The development of a taxonomy is a common mechanism to unify and formalise information scattered in the literature~\cite{Kwasnik99}. In particular, taxonomies in software engineering have been proposed with the aim of formalising current research practices~\cite{Felderer14}. Taxonomies like SERP (Software Engineering Research and Practice) are also a good instrument to support academia-industry communication~\cite{Engstrom17}, clearly defining the problem under study and mapping challenges to solutions~\cite{Petersen14}. In fact, the lack of communication and differences in terminology have been identified as key factors hampering the transfer of RT research to industry~\cite{Minhas20}. In the context of software testing, SERP-test ---an adaptation of SERP to this area--- allows structuring software testing research with respect to testing tasks, context constraints and expected effects. Other relevant taxonomies in the area cover existing techniques~\cite{Saeed16} and tools~\cite{Costa20}. One taxonomy also analysed the alignment of testing and requirements activities based on information transfer~\cite{Unterkalmsteiner14}. Recently, Pan et al. have reviewed ML-based TCP literature using four dimensions of classification (techniques, features, evaluation metrics and reproducibility)~\cite{Pan21}. As they mention, such dimensions can be viewed as a form of taxonomy to help classify new TCP proposals, although they do not provide a thorough characterisation of the features as we do with respect to information attributes.

A taxonomy providing full characterisation of information attributes based on data acquisition and measurement would allow researchers and practitioners to increase awareness when conceiving TCP methods under different scenarios of information availability and maintenance costs. With this aim, this paper poses a main research question (RQ1), about our taxonomy, and two additional ones based on its use:

\begin{itemize}
    \item RQ1: What information attributes have been defined for TCP and how can they be characterised?
    \item RQ2: What information attributes are required and used for the application of TCP in the industry?
    \item RQ3: What information attributes have been considered in ML-based TCP methods and to what extent are they aligned with those used in industry-oriented studies?
\end{itemize}

To answer RQ1, we propose \acronym, a taxonomy of Test case Prioritisation Information Attributes. \acronym offers a common terminology to describe and classify the information attributes currently supported by TCP methods, providing both researchers and practitioners a unified framework to analyse the alternatives from different perspectives. This can also help guide future evolution of TCP methods as well as in finding new information attributes that have not been considered. The development of the \acronym taxonomy is guided by the analysis of more than 100 research papers, from which 91 different attributes are defined and evaluated in terms of scope, collection and measurement. Supported by the knowledge derived from the taxonomy, we then try to shed light on RQ2 by focusing on the information attributes in the 23 papers from our literature search reporting some kind of industrial evaluation. More specifically, our study analyses the information attributes more frequently used and how they are combined in TCP methods. Similarly, 34 ML-based TCP methods are inspected to discover the preferred information attributes to build data features. In addition, we study the relationship between information attributes and types of learning approaches. We contrast the results from both analyses to discuss RQ3 and propose steps forward. Finally, an initial evaluation of the taxonomy is conducted based on TCP approaches proposed in the recent literature and informal interviews with software quality assurance (QA) professionals from three companies.

Our analysis reveals that less than half of the information attributes defined in the \acronym taxonomy have been applied for TCP in the context of academia-industry collaboration, whose methods are guided by a combination of no more than five attributes. Also, several unique information attributes have appeared in the industrial studies only, and those related to changes, faults and risks were highlighted as the more important during our industrial interviews. A larger variety of attributes and combinations of information sources is found in the ML literature. It does not necessarily mean that all ML-based methods are making unrealistic assumptions about information availability and relevance, since some of them discuss these factors as part of their industrial evaluation. However, as the interest in the field grows, so does the risk of not properly directing the potential of ML towards solving those problems that industry really demands, and towards information that is cost-effective to collect and maintain the quality of. In this sense, the \acronym taxonomy aims to be a reference for the design of cost-effective ML-based prioritisation methods with respect to the data features used for learning. With this aim, the taxonomy is available as a GitHub repository open to contributions from the testing community (see additional material in page~\pageref{sec:additional}). 

The rest of the paper is organised as follows. Section~\ref{sec:background} presents background on TCP and ML applied to RT. Section~\ref{sec:related} compiles previous work presenting a summary of industry-focused papers addressing the TCP problem, as well as information-related classifications relevant for RT and ML. Section~\ref{sec:methodology} describes the methodology followed for the taxonomy construction, according to a bottom-up approach. The \acronym taxonomy is presented in Section~\ref{sec:taxonomy}, including its evaluation. We then analyse the information attributes from an industrial perspective (Section~\ref{sec:rq2_analysis}) and discuss ML applicability (Section~\ref{sec:rq3_analysis}). Based on the taxonomy and the literature analysis, Section~\ref{sec:discussion} provides a discussion about ML-based TCP oriented to industry. The threats to validity are enumerated in Section~\ref{sec:threats}. Finally, Section~\ref{sec:conclusion} highlights some concluding remarks.

%--------------------------------
% 2 - BACKGROUND
%--------------------------------
\section{Background}\label{sec:background}

This section introduces the most relevant concepts and terminology to this paper. Firstly, test case prioritisation is introduced, describing the common information sources. Secondly, the role of ML on RT is addressed in detail, explaining the approaches currently explored.

\subsection{Test case prioritisation}\label{subsec:tcp}

During RT, three main activities are carried out, namely test suite minimisation, test case selection and test case prioritisation~\cite{Yoo12}. Test case prioritisation seeks to rank test cases so that more relevant test cases, \eg those associated to high fault detection rate, are executed first. In their literature review, Khatibsyarbini et al. find 19 different methodologies to address the prioritisation problem~\cite{Khatibsyarbini18}, from which the authors derive a general flow for TCP comprised of four steps: 1) \emph{data preparation}, which consists in identifying the information to be used, \eg requirements, models or code; 2) \emph{data measurement}, which usually implies metric computation or dependency analysis; 3) \emph{prioritisation}, which is the actual execution of the technique using the information obtained from the previous step; and 4) \emph{monitorisation}, which evaluates the adequacy of the TCP method in terms of fault detection or other criteria.

Focusing on the first two steps, input information might come from a variety of sources. The structural coverage represents one of the first and probably most used information attributes for white-box TCP~\cite{Rot01}. Reaching high coverage has been traditionally considered as a good indicator of testing effectiveness~\cite{Lyu94}, and several measures have been focused on different code constructs (branches, statements, methods, etc.) However, some studies have empirically analysed the relation between coverage and the fault detection capability of test cases~\cite{Cai05,Hemmati15}, showing that their effectiveness depends on the choice of the coverage measure, how it is combined and the particularities of the test cases. Besides, code coverage information might be difficult to extract and keep updated for certain systems~\cite{Luo19}. As a consequence, many authors have turned to black-box approaches ---those not requiring access to the SUT code--- or grey-box proposals ---those solely relying on test code---~\cite{Henard16}, thus opening up the spectrum of information explored so far. Kim and Porter were the first to introduce the history of test executions in a prioritisation probabilistic model~\cite{Kim02}. Similarly, Fazlalizadeh et al.~\cite{Faz09} define the priority of a test case based on its past effectiveness, \ie whether it revealed a fault in previous testing sessions. Estimations of fault-exposing potential of test cases have been investigated too~\cite{Elb02}, though it requires mapping test cases to faults a priori, \eg via mutation testing. Information extracted from code changes~\cite{Elb03}, risk analysis~\cite{Wan18} and test case similarity~\cite{Led12} represent other examples of inputs for TCP.

\subsection{Machine learning for regression testing}\label{subsec:ml-testing}

Applications of ML for software testing can be traced back to the early 90s, with studies focused on test effort estimation~\cite{Briand92,Cheatham95} and test case generation~\cite{Anderson92,Bergadano93}. A recent mapping study gives an overview of how the field has evolved over time and which other testing activities have been tackled from a ML perspective~\cite{Durelli19}. Test case prioritisation appears as the matter of study of five papers~\cite{Ton06,Che11,Len13,Bus16,Spi17}. Recently, the application of ML to TCP has been considered significant enough to deserve its own category when classifying TCP techniques~\cite{CastroCabrera20}.

ML algorithms work with a set of instances, for which a collection of data features is provided. In software testing, data can be obtained by preprocessing testing and/or SUT code, \eg linking test cases to the entities they cover, or collecting information about the testing process and its outcomes, \eg whether test cases failed or not in previous testing cycles. Usually, the raw data extracted has to be preprocessed by applying methods to remove noise, deal with missing values, scale and discretise data values, or select a subset of features and instances. Preprocessing is known to be a fundamental step that can greatly improve the results of ML in software engineering~\cite{Huang15}. Once data is prepared, the ML algorithm automatically analyses it to discover hidden knowledge to build decision models able to make predictions, describe patterns or provide recommendations~\cite{Durelli19}. Depending on the characteristics of the data and the specific ML goal, different ML approaches can be found~\cite{Zhang05}:

\begin{itemize}
\item \emph{Supervised learning}. These approaches try to predict a target variable from the data features, requiring a labelled dataset for training in which each instance is associated to an output value. Regression and classification are the most frequent techniques, focused on predicting a continuous variable or a class value, respectively.
\item \emph{Unsupervised learning}. These methods have a descriptive nature, not requiring labels in the dataset. Clustering groups instances according to a similarity measure which is computed from the features. Pattern mining and association rule mining infers patterns and rules, respectively, describing the characteristics and data relations within the dataset. Topic modelling combines natural language processing and text mining to infer frequent topics from text. 
\item \emph{Semi-supervised learning}. It is possible to train with a low percentage of manually labelled instances and predict the class of the remaining instances. Labels can be obtained from different ways, including manual processes and active learning.
\item \emph{Reinforcement learning}. The idea is to automatically learn a reward function, which reinforces those actions taken by the algorithm with positive results. The adaptive process might be automatically guided by a supervised learning algorithm, \eg neural networks, or founded on fixed rules defined a priori.
\item \emph{Deep learning}. These advanced models are comprised on multiple processing layers that learn representations of the data at different level of abstraction, automatically adapting its internal structure. Deep models are becoming increasingly popular due to its ability to deal with large sets of unstructured data.
\end{itemize}

%--------------------------------
% 3 - RELATED WORK
%--------------------------------

\section{Related work}\label{sec:related}

This section presents related work divided into two topics: (1) industrial studies in regression testing, with emphasis on the difficulties that arise when transferring prioritisation techniques to industrial environments, and (2) previous works proposing taxonomies and other classifications for information sources in software testing.

\subsection{Industry studies in regression testing}\label{subsec:related-industry}

In 2012, Yoo and Harman pointed out the limited evaluation and application of RT techniques in industrial environments~\cite{Yoo12}. Goals pursued by RT, such as fault detection or test case dependency analysis, are indeed essential for the software industry, but they are not so easily addressed when large systems coexist~\cite{Onoma98} or RT is subject to time~\cite{Mar13} and resource constraints~\cite{Wan16}. Besides, most of RT research operates at unit testing level, but industry also demands RT methods either to work at the system testing level~\cite{Ule18} or to be targeted to acceptance testing~\cite{Li13}. Aspects related to the system, such as the size, complexity, type and heterogeneity of the SUT, have been identified as important factors to put RT into practice~\cite{Ali19EMSE}. The number and magnitude of systems imply that RT activities might take several days~\cite{Car11} or even weeks~\cite{Ala18}, but changes are committed at a higher frequency~\cite{Mar13}. 

The adoption of RT also presents severe limitations if the \emph{information sources} required by a particular technique are unavailable~\cite{Ali19EMSE} or are not aligned with the company needs. Furthermore, the relevance of the information attributes might depend on organisational practices and project-specific factors~\cite{Kan17,Minhas20}. Engstr\"om et al. conclude that a history-based approach better fits the interests of their industrial partner, since detection effectiveness, test case age and time since last execution are highly significant~\cite{Eng11}. In another study, the industrial collaborators consider fault probability, time and cost associated to test case setup and execution, and requirement coverage as the most relevant information~\cite{Tah16}. Change-related attributes were identified as the primary information for testing an embedded system~\cite{Minhas20}.

Data availability is not enough, since the cost of collecting, storing and updating it can be prohibitive~\cite{Zha09-icsm}. From cases studies, it is observed that RT in industry is still a manual process and supporting tools for extracting data from the testing process are not always available~\cite{DiN15}. Even if test management systems are considered, the time and effort required to extract the information should not be disregarded~\cite{Li13}. To cope with this, some authors update data asynchronously~\cite{Bus16} and build specific tools to automate the extraction of information~\cite{Lac16,Ule18}.

\subsection{Other classifications of information sources for software testing}{\label{subsec:related-classifications}

Two studies analyse information sources for TCP, establishing some classifications. Firstly, Sampath et al. propose a formal definition of TCP methods that combine information sources, referred as hybrid approaches~\cite{Sampath13}. They identify three alternatives: 1) \emph{rank}, which establishes an order of application; 2) \emph{merge}, which simultaneously uses the available information; and 3) \emph{choice}, which selects one attribute among equally important factors. As part of their analysis, they identify 44 TCP papers applying these approaches, only providing the list of the information attributes used in each case. Information sources are also analysed in a recent mapping study of 35 TCP techniques for continuous integration (CI)~\cite{PradoLima20}. Nine categories of sources are defined to classify TCP methods, though a list of information attributes for each one is not provided. According to the findings, failure and execution histories are preferred over coverage, and 60\% of the studies combine two or three sources. Compared to these works, our taxonomy provides a formal definition of input information for TCP at the attribute level, providing a deeper analysis of their characteristics from three dimensions. Also, \acronym is built from a wider collection of publications, as it is not restricted to the way information attributes are combined or the scope of CI testing.

Focusing on ML-based testing approaches, Durelli et al. analyse them with respect to the data features used for learning~\cite{Durelli19}. The majority of proposals extract information from the test cases, but it is frequently used in combination with test suite metrics and source code metrics. In this line, Noorian et al. propose a more formal framework to classify ML research on software testing~\cite{Noorian11}. Divided into five dimensions, the framework serves to categorise both the testing task and the ML technique. In terms of data features, referred as \emph{elements to be learned}, the authors distinguish among software metrics, software specification, control/call graphs, test cases, execution data, failure reports and coverage. A small collection of studies ---two out of the four analysed papers deal with RT--- was considered to validate the framework. Recently, Pan et al. have conducted a systematic literature review of test case selection and prioritisation using ML~\cite{Pan21}. As part of their literature analysis, one RQ focused on the types of data features used for training, which are classified into five groups: code complexity, textual data, coverage information, user inputs and history. For each group, the authors provide a description and examples of data features appearing in the 29 papers under analysis. As part of the discussion of findings, they argue that a more detailed study of the most relevant features and tools to extract their values is an interesting future research direction to evaluate the cost-benefit of ML approaches. All these works cover ML methods applied to any testing task, not carrying out a specific analysis on how information for TCP is extracted and combined. Since our taxonomy is focused on information attributes for any TCP method, it does not include specific dimensions to categorise the techniques as Noorian et al. or Pan et al. do. However, we include an analysis of ML-based TCP techniques to understand the relationship between information attributes and the learning task.

Our work is also related to studies reflecting on industry-oriented research. Recently, Ali et al. have analysed the industrial relevance of RT research~\cite{Ali19EMSE}. These authors propose three taxonomies to formalise the context, effect and information factors influencing the industrial applicability of RT. The information taxonomy only contains two levels: the first level establishes nine entities (requirements, design artefacts, source code, intermediate code, binary code, test cases, test execution, test reports and issues), while the second level enumerates 50 related attributes in total. Being focused on 38 industrial case studies, the list of attributes is considerable lower than the one covered in our taxonomy, and some of them might have not been conceived for TCP. Finally, our taxonomy includes additional dimensions to fully characterise the information attributes, supporting further analysis of the factors affecting data preparation for TCP.

Finally, literature reviews for other testing activities also include information-related classifications. For instance, a taxonomy for model-based testing defines a dimension of test selection criteria~\cite{Utting12}, which includes coverage and fault-based metrics. More recently, a literature review on test case generation for agent-based models analyses the information artefacts from which test cases are derived~\cite{Clark21}. Related to ML, a survey on software fault prediction classifies the selected studies depending on the type of code metrics used as data features~\cite{Pandey21}.

%--------------------------------
% 4 - METHODOLOGY 
%--------------------------------
\section{Methodology}\label{sec:methodology}

This section describes the method applied to develop our taxonomy. Traditionally, taxonomies are built following a top-down or a bottom-up approach~\cite{Broughton15}. In top-down approaches, the categories of the taxonomy are defined on the basis of previous classifications and formal concepts. In contrast, bottom-up approaches are founded on the analysis of existing works in the area of interest with the aim of discovering patterns and identifying the underlying concepts~\cite{Unterkalmsteiner14}.
\acronym had to be built by following a bottom-up process, since there is no classification of information sources or their attributes for RT yet. Furthermore, the application of ML for TCP is gaining increasing attention in the last years, meaning that new techniques are continuously emerging, and their information attributes have not been covered by any secondary study yet. The rest of this section explains in detail the phases and activities carried out to build \acronym. Next, the literature search process to collect relevant studies is described. 

\subsection{Building the taxonomy}\label{subsec:process}

Our process to plan, design and build the taxonomy is highly aligned with the phases proposed by Bayona-Or\'e et al.~\cite{BayonaOre14}. Their method consists of the following phases: 1) \emph{planning}, 2) \emph{identification and extraction of information}, 3) \emph{design and construction of the taxonomy}, 4) \emph{testing and validation} and 5) \emph{deployment of the taxonomy}. Each phase is comprised of a sequence of activities, which were derived after reviewing existing methods and common guidelines for developing taxonomies. These general activities are then adapted to the particular application domain.

The first phase, \emph{planning}, starts with the identification of the area of study for which our taxonomy, \acronym, is conceived, as well as the definition of its objectives. Particularly, \acronym is framed within the knowledge area of ``software testing''~\cite{Bourque14} and, more specifically, RT~\cite{Yoo12}. The main goal of the \acronym taxonomy is to formalise and structure the testing information attributes often used in TCP, and how they are mapped into data features used by prioritisation techniques based on ML. The resulting taxonomy seeks to establish a common terminology, to identify relevant attributes for prioritisation from an industrial perspective, to find evidences of how they are combined, and to serve as a basis to discuss current challenges in the adoption of ML-based prioritisation techniques in industrial environments. In terms of scope, the taxonomy is not intended to cover the full spectrum of prioritisation techniques available in the literature, but to synthesise the most relevant approaches with respect to the testing inputs. Only ML-based techniques, especially the ones that have been used in industry, are analysed in depth in order to understand how such information is transformed into data features to feed either descriptive or predictive algorithms. The \acronym taxonomy responds to the needs of both practitioners and researchers with interest in the application of ML methods for testing large-scale, typically industrial, systems. The taxonomy provides them with practical knowledge for the data preparation phase, with special emphasis on constraints that might hamper the extraction and maintenance of the information attributes by companies. In this sense, the team responsible for developing the taxonomy is comprised of researchers with experience on the practicalities of ML in different domains and specific background on the application of AI to software engineering. Also, two authors have previously worked in industry, and one of them currently works in close collaboration with industry, providing them with AI- and ML-based methods for testing their software systems.

The second phase consists in the \emph{identification and extraction of information}. Firstly, the number and type of literature sources is determined. Given that our taxonomy is developed using a bottom-up approach, a literature search of prioritisation methods has to be conducted. The search process combines manual inspection of well-known secondary studies focused on different perspectives of RT~\cite{Yoo12,Khatibsyarbini18,Durelli19} with automatic search to ensure that recent proposals are included. Details on the search process and the selection of relevant studies are separately reported in Section~\ref{subsec:search}. The classification schemes applied in secondary studies~\cite{Minhas20}, together with existing software testing taxonomies~\cite{Saeed16,Ali19EMSE}, also constitute a source of concepts and terms to derive candidate names for categories of the \acronym taxonomy. Case studies and industrial experiences reported in the RT literature also serve to find evidence of the difficulties surrounding the extraction of information attributes, as well as to analyse their relevance in real-world contexts.

The third phase involves the \emph{design and construction of the taxonomy}. Firstly, the list of information attributes is extracted from the selected studies. During this process, new values to classify each work are included as needed, and decisions are made about how similar ideas could be abstracted to more general concepts. For the \acronym taxonomy, such decisions consisted in identifying attributes measuring the same concept and agreeing a common definition. When two works differ in the extraction process or the data representation, the differences are annotated and multiple values are assigned to the corresponding categories. Once the extraction process is completed, the first level of the taxonomy is set by grouping those categories that are related. As for terminological decisions, \eg what name is finally assigned to each attribute, we analyse naming rules frequently adopted in the literature ---specially in other taxonomies--- and how the concept is usually referred in industry. Historical reasons also influence the decision, \ie the name adopted by the first work using the information attribute prevails when no common name is found in subsequent studies. The construction of the taxonomy finishes with the consolidation of the second level of categories. 

The fourth phase refers to \emph{testing and validating} the proposed taxonomy. A first validation should focus on the ability of the \acronym taxonomy to capture the diversity of information attributes used by current prioritisation techniques. A sample of research papers published after the literature search period could serve to discover missing information not covered by the taxonomy due to its bottom-up building process. Such an analysis and its conclusions are discussed in Section~\ref{subsec:eval}. A closer look into already published ML studies can provide evidence on the effectiveness of the \acronym taxonomy at this stage, including an analysis of data features used in combination and a discussion on how well they are aligned with the industrial needs reported in case studies. In addition, we conducted informal interviews with five software QA professionals from three companies, two located in Spain and one located in Sweden. These companies (a consulting services company, a product-oriented start-up, and a global technology company working in the aerospace domain) work on different domains and under diverse testing conditions. These interviews were planned with the aim of contrasting the information extracted from the literature with current RT industrial experiences, as well as getting feedback on the potential use of our taxonomy in their context.

Finally, we decide not to elaborate a precise plan for the \emph{deployment of the taxonomy}, since its target audience are other researchers and practitioners who will find the necessary information in this paper and its additional material (see page~\pageref{sec:additional}). In particular, the taxonomy is openly available in a dedicated repository to allow knowledge access and help future deployment. As for its maintenance, the TePIA taxonomy is a live artefact, so it is expected to evolve as new TCP methods appear. This could imply adding attributes or extending the values of some characteristics to cover new ideas. New versions of the taxonomy will be created by periodically revising the TCP literature. The public repository allows keeping the taxonomy up-to-date, controlling its versions, as well as planning new releases as we receive comments and contributions from the TCP community.

\subsection{Literature search}\label{subsec:search}

When building a taxonomy using a bottom-up approach, the literature search process should ensure that a representative sample of studies is selected. Some authors follow well-known guidelines for systematic literature reviews to retrieve relevant studies~\cite{Felderer14, Saeed16, Costa20}. Other authors have opted for a mixed approach~\cite{Unterkalmsteiner14,Ali19EMSE}, starting with a list of references collected from previous studies that are later complemented with snowballing procedures or additional searches in databases. Considering that advances on RT have been systematically reviewed multiple times in the last decade~\cite{Khatibsyarbini18}, their list of selected primary studies serves us to obtain a ``golden set'' of prioritisation methods. References from two recent papers, a mapping study of ML applied to software testing~\cite{Durelli19} and literature review on industry-relevant RT~\cite{Ali19EMSE}, are also considered since they focus on key aspects for our taxonomy. 

After identifying candidate papers from these secondary studies, we proceed to decide whether they are relevant for our taxonomy. The scope of our analysis is TCP methods proposed for testing implemented software systems, with special focus on methods using ML techniques. The following exclusion criteria are applied first:

\begin{enumerate}
    \item The content of the paper is not accessible.
    \item The paper describing the method is not written in English.
    \item TCP papers not based on ML techniques that are retrieved from automatic search are not published in reference conference and journals (details are given below).
\end{enumerate}

For the remaining papers, inclusion criteria are considered to confirm that the proposed method is within scope:

\begin{enumerate}
    \item The paper proposes or applies a prioritisation method for test cases.
    \item The paper describes the sources from which information for TCP is extracted.
    \item Inputs to the TCP methods are directly collected from the SUT or its testing process, \ie methods based on requirement information only or founded on model-based testing are not considered.
    \item The method is not tied to a particular type of system, such as product lines and mobile applications, whose characteristics make the TCP method not applicable to other SUT.
\end{enumerate}

Next, we detail the number of papers available in and selected from each secondary study: 

\begin{itemize}
    \item The survey of RT by Yoo and Harman~\cite{Yoo12} contains a list of 159 studies published between 1977 and 2009, from which 47 corresponds to test case prioritisation. After applying our selection criteria, we keep 30 papers. One reports industrial evaluation and three use ML, while the rest of papers apply other techniques, \eg search, over benchmarks, \eg SIR repository.

    \item The systematic literature review of test case prioritisation by Khatibsyarbini et al.~\cite{Khatibsyarbini18} analyses 69 papers, whose publication date ranges from 1999 to 2016. 41 papers were considered in scope, five are industrial case studies and three papers apply ML. Only eight papers (one of them using ML) coincide with papers selected from the previous survey.

    \item The mapping study of ML applied to software testing by Durelli et al.~\cite{Durelli19} summarises 48 papers published up to mid 2018. Five papers ---two using industrial data--- propose ML-based prioritisation techniques, so they all are included here. Four of these papers ---two are dated after 2016--- were not found in previous steps.
    
    \item The systematic literature review conducted by Ali et al.~\cite{Ali19EMSE} as part of their look for industry-relevant RT research provides us a list of 38 papers. The search was carried out in 2016, and included an initial search based on the list provided by Yoo and Harman~\cite{Yoo12}. We keep ten of the 38 papers (all reporting case studies), one of them applying ML. Two and one paper(s) can be found in the RT survey~\cite{Yoo12} and review~\cite{Khatibsyarbini18}, respectively.
\end{itemize}

Since the previous studies cover literature published up to 2018, and only a few ML studies were found, recent proposals could be missing. Therefore, we follow a snowballing procedure and perform additional searches on a scientific database. As result of snowballing, 11 additional papers are considered relevant for our taxonomy (two are industrial and nine use ML). For the automatic search, we choose Scopus because it retrieves conference proceedings as well as journal articles from diverse publishers. Three search strings using general terms were executed by March 2020 (see Table~\ref{tab:search}), but restricted to publications up to (and including) 2019. Research works published in 2020 are left for taxonomy evaluation (see Section~\ref{subsec:eval}), and were retrieved by December 2020. The first two strings in Table~\ref{tab:search} allow detecting any mention to ML linked to TCP or, more generally, RT. By running these strings, we ensure that ML works that might have been omitted by secondary studies are included here. The third search string seeks for any publication on RT, and focused on prioritisation, published between 2017 and 2019, thus complementing the list of general TCP methods obtained from the reviews previously published up to 2016~\cite{Ali19EMSE,Khatibsyarbini18}. In this case, the number of results is considerably higher, and a first screening of papers shows redundancy in the information and methods applied for TCP with respect to already selected papers. Therefore, we only keep the works published in top conferences and reference journals that might present the most significant advances in the area. More specifically, we select SE conferences (ICSE, ESEC/FSE, ICST and ISSTA) appearing in the CORE ranking. As for journals, they should be JCR-indexed in the SE category. The only exceptions are journal papers that propose a ML-based method, which can be published in any other JCR category.
Table~\ref{tab:search} details the number of papers found and selected for each search string. From a total of 33 distinct papers, we found 10 industrial studies and 19 papers applying ML. Three appeared on literature reviews and four coincide with papers found via snowballing. Figure~\ref{fig:references} summarises the source of the 110 papers that compose the reference list. Numbers in brackets indicate the number of papers using ML plus those based on any other technique. All the papers lying in the intersections are ML approaches.

\begin{table*}[ht]
  \caption{Search strings used to find relevant RT studies and number of results.}
  \label{tab:search}
  \begin{tabular}{clcc}
    \toprule
    &\textbf{Search string} & \textbf{Results} & \textbf{Selected}\\
    \midrule
    \#1&\texttt{``regression testing'' and ``machine learning''}  & 27 & 10\\
    \#2&\texttt{``test case prioritization'' and ``machine learning''} & 24 & 9\\
    \#3&\texttt{``regression testing'' and ``prioritization''}  & 177 & 19\\
    \midrule
    & \multicolumn{2}{l}{\textbf{Total distinct papers}}&\textbf{33}\\
    \bottomrule
  \end{tabular}
\end{table*}

\begin{figure}[ht]
  \centering
  \includegraphics[width=0.4\linewidth]{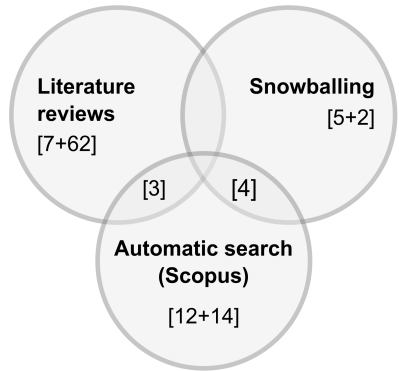}
  \caption{Distribution of references used to build the taxonomy. }
  \label{fig:references}
  \Description{The figure shows how many papers were found from each of the sources described in the text, as well as those papers appearing in more than one source.}
\end{figure}

%--------------------------------
% 5 - TAXONOMY
%--------------------------------
\section{The \acronym taxonomy}
\label{sec:taxonomy}

This section presents the \acronym taxonomy. Firstly, Section~\ref{subsec:overview} explains its dimensions and categories. Next, Sections~\ref{subsec:group1} to~\ref{subsec:group3} analyse the most relevant characteristics of the information attributes. For each attribute, the reference where it was first defined is included. The interested reader is referred to the additional material for a more in-depth analysis and the full list of references using each attribute. Finally, Section~\ref{subsec:eval} presents the evaluation of the proposed taxonomy.

\subsection{Overview}\label{subsec:overview}

The \acronym taxonomy formalises the definition of information attributes for TCP. It is organised into dimensions, which are further decomposed into categories. One or more values per category are assigned to describe the identified attributes. The taxonomy thus gives a structured way to describe and group information attributes. Figure~\ref{fig:taxonomy} shows the dimensions and categories of the \acronym taxonomy, which are described next, together with their possible values. Then we present the actual attributes we found in the literature based on how they map into the taxonomy.

\begin{figure}[ht]
  \centering
  \includegraphics[width=\linewidth]{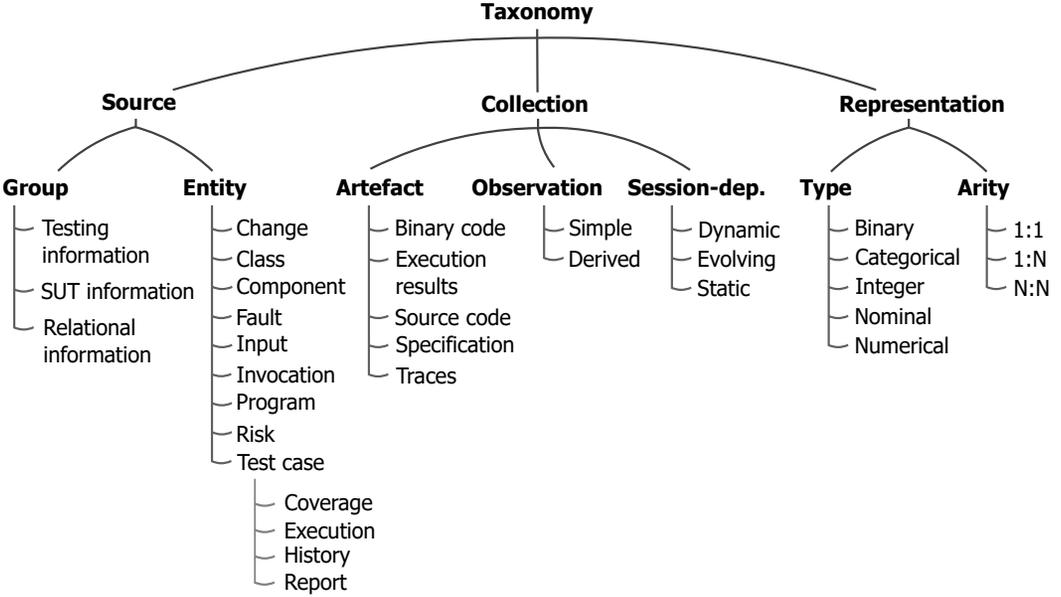}
  \caption{Taxonomy of information attributes for test case prioritisation.}
  \label{fig:taxonomy}
  \Description{A tree depicting the hierarchical decomposition of the \acronym taxonomy dimensions and categories.}
\end{figure}

\begin{enumerate}
    \item \textbf{Source}. This dimension specifies where the information is generated or originates. It is further decomposed into the following categories:
    
    \begin{enumerate}

        \item \emph{Group}. A high-level view of the nature of the information, divided into three categories. Firstly, \textit{testing information} attributes only require the analysis of test cases or its outcomes. Secondly, \textit{SUT information} attributes are based on characteristics of the system under test (SUT). A third category, referred as \textit{relational information}, lies in between with the aim of grouping those attributes that need to establish connections between test cases and the SUT, \eg which functionalities are covered by each test case.
        
        \item \emph{Entity}. An abstract representation of the type of element from which the information is going to be extracted. The element can be a part of the system at different levels of abstraction (program, component or class). It can also represent an action generating information about the SUT, such as an invocation or change, or something derived from the specification, \eg risk and inputs. Focusing on the testing activity, test cases and the faults they expose are important entities too. Test case, as a primary entity of the TCP process, is further decomposed to analyse the different aspects considered so far.
    \end{enumerate}

    \item \textbf{Collection}. This dimension serves to analyse how the information attribute is obtained from the entity. The following categories are defined:
    \begin{enumerate}
        \item \emph{Artefact}. The artefact that contains the entity and should be parsed or instrumented to obtain the attribute value. The possible options are: binary code, execution results, source code, specification (either of the test case or SUT), and traces.
    
        \item \emph{Observation}. A category to specify whether the attribute value is directly obtained as a raw value from the artefact or if it is derived from other measures.

        \item \emph{Session-dependent}. This category specifies whether the attribute value changes when a new testing session is carried out (dynamic) or not (static). A third option, named ``evolving'', indicates that changes in the SUT might affect testing outcomes.
    \end{enumerate}

    \item \textbf{Representation}. This dimension seeks to describe how the information attribute is computationally stored after processing the collected data. Two categories are defined:

    \begin{enumerate}
    \item \emph{Type}. The sort of computational variable that is assigned to the attribute. The nomenclature by Pyle~\cite{Pyle99} is considered: binary, categorical, integer, nominal ---including text--- and numerical (floating-point number).
    
    \item \emph{Arity}. Represented as a tuple N:N, it describes whether several entities are involved in data computation. The first element specifies if the attribute is computed for a single entity (\eg a test case) or a group of entities (\eg a test suite). The second element indicates whether the value depends on a single entity (\eg the test case itself) or many (\eg other test cases o several pieces of code). In short, this category will allow establishing one-to-one, one-to-many and many-to-many relations among the values needed to compute the attribute.
    \end{enumerate}
\end{enumerate}

These dimensions and categories were consolidated after successive iterations. During the process, a total of 91 information attributes found in the literature were classified. Notice that values for each category were assigned considering what has been reported in the literature, \ie how authors define and use the attributes. However, other values might be applicable in other contexts, as will be pointed out in particular cases.

\subsection{Testing information}\label{subsec:group1}

\begin{table*}[ht]
\caption{ Classification of attributes based on testing information.}
\label{tab:eval-test-summary}
\scalebox{0.9}{
\begin{tabular}{lllllll}
\toprule
\multicolumn{2}{l}{\textbf{Source}} & \multicolumn{3}{l}{\textbf{Collection}} & \multicolumn{2}{l}{\textbf{Representation}} \\
\toprule
\emph{Entity} & \emph{Attribute} & \emph{Artefact} & \emph{Observation} & \emph{Session} & \emph{Type} & \emph{Arity} \\
\midrule
Test case
& Age                   & Exec.  & Simple & Evolving      & Num.  & 1:1  \\
property 
& Estimated fault detection       & Spec.  & Simple & Static        & Cat.  & 1:1  \\
& Resources             & Spec.  & Simple  & Static       & Bin.  & 1:N  \\
& Size                  & SC     & Simple  & Static       & Num.  & 1:1  \\
& Static priority       & Spec.  & Simple  & Static       & Num.  & 1:1  \\ 
& Status                & Spec.  & Simple  & Dynamic      & Bin.  & 1:1  \\ 
& Textual description   & Spec.  & Derived & Static       & Num.  & 1:N  \\ 
& Type of test          & Spec.  & Simple  & Static       & Bin.  & 1:1  \\ 
\midrule
Test case
& Allocation time       & Exec.  & Simple  & Static       & Num.  & 1:1  \\
execution 
& Cost                  & Spec.  & Simple  & Static       & Cat.  & 1:1  \\
& Execution time        & Exec.  & Simple  & Dynamic      & Num.  & 1:1  \\
& Resource utilisation  & Traces & Derived & Dynamic      & Num.  & 1:N  \\
& Total time            & Spec.  & Simple  & Static       & Cat.  & 1:1  \\
\midrule
Test case                 
& Effectiveness             & Exec. & Simple & Dynamic      & Bin. & 1:1  \\
report 
& Failure frequency         & Exec. & Simple & Dynamic      & Num.  & 1:N  \\
\midrule
Test case
& Historical effectiveness  & Exec.  & Derived  & Dynamic   & Num.  & 1:N  \\
history 
& Historical executions     & Exec.  & Derived  & Dynamic   & Int.  & 1:N  \\
& Historical fails          & Exec.  & Simple   & Dynamic   & Int.  & 1:N  \\
& Historical verdicts       & Exec.  & Simple   & Dynamic   & Bin.  & 1:N  \\
& Previous execution        & Exec.  & Simple   & Dynamic   & Bin.  & 1:1  \\
& Previous priority         & Exec.  & Simple   & Dynamic   & Int.  & 1:1  \\
\midrule
Test case
& Implementation dependencies     & Spec.  & Derived & Static       & Int.  & 1:N  \\
dependency
& Joint execution       & Spec.  & Simple  & Dynamic      & Cat.  & N:N  \\
& Verdict pattern       & Exec.  & Derived & Dynamic      & Nom. & 1:N  \\
\midrule
Test case 
& Code similarity           & SC/Tra. & Derived & Static  & Num.  & N:N  \\
similarity 
& Input similarity          & Spec. & Derived & Static      & Num.  & N:N  \\
& Text similarity           & Spec. & Derived & Static      & Num.  & N:N  \\
\bottomrule
\end{tabular}
}
\begin{tablenotes}
\item \footnotesize{Artefact: Exec=Execution results, SC=Source code, Spec=Specification, Tra=Traces}
\item \footnotesize{Type: Bin=Binary, Cat=Categorical, Int=Integer, Nom=Nominal, Num=Numerical}
\end{tablenotes}
\end{table*}

Table~\ref{tab:eval-test-summary} shows the attributes belonging to (source) group \textit{testing information}. Although only a single entity, the test case, has been identified for this group, a wide variety of aspects can be measured. Five attributes represent test case properties obtained as simple observations: the \textit{size}, usually measured as lines of code~\cite{Noo17}; the \textit{resources} required to run the test case, with arity 1:N because it is expressed as a list~\cite{Wan16}; the \textit{fault-detection} capability~\cite{Tah16} and the \textit{static priority}~\cite{Eng11}, both being estimated by a tester; and the \textit{type of test}, in binary form to distinguish acceptance tests from functional tests~\cite{Kan17}. We note that the latter attribute might include other types of test cases meaningful in other testing scenarios, such as system, integration or performance tests, changing its type to categorical. Two other properties change as long as the test case is executed, namely its \textit{age}~\cite{Eng11} (time since its creation) and its \textit{status}~\cite{Noo17} (whether it is new or modified). Similar to the \textit{type of test}, the \textit{status} of the test case could need a more fine-grained classification and become a categorical attribute. The \textit{textual description} of the test case is the only attribute derived from the test case definition, as it is expressed as a bag of words or topics whose frequency is automatically determined to get a numerical score~\cite{Lac16}. 

The rest of the attributes refer to properties of the test cases collected during or after their execution. From test case execution, TCP techniques might use attributes such as \textit{allocation time}~\cite{Wan16}, \ie time needed to prepare or configure the test case, estimated \textit{cost} of implementation and setup~\cite{Li13}, \textit{execution time}~\cite{Sri02}, \textit{resource utilisation}, \ie CPU, memory and I/O required by the test case~\cite{Bha18}, and \textit{total time}, \ie time needed to implement, configure and run~\cite{Tah16}. Once a test session finishes, the test case \textit{effectiveness}~\cite{Kim02}, \ie whether it passes or fails, and the \textit{failure frequency}~\cite{Ala18}, \ie ratio of fail verdicts, can be measured from the test case report. Although most of these attributes are numerical in nature, cost and time have been expressed using categorical variables too, suggesting that the specific value is not so relevant and is rather classified into broad groups.

Outcomes from previous executions, known as test case history, allow including a long-term view of TCP performance. The following attributes, which have a dynamic nature, have currently been applied: \textit{historical executions}, which counts the number of times a test case has been executed~\cite{Faz09}; \textit{historical effectiveness}, defined as the ratio between test case fails and runs~\cite{Faz09}; \textit{historical fails}, which refers to the number of sessions for which the test case has failed~\cite{Noo17}; \textit{historical verdicts}, \ie the sequence of results in the last $n$ sessions~\cite{Mar13}; \textit{previous execution}\cite{Kim02} and \textit{previous priority}~\cite{Lin13} indicate whether the test case has been executed and its ordering position in the previous section, respectively. Only the \textit{historical effectiveness} and the \textit{historical executions} require some previous calculations, \ie they are derived. The former is defined as a ratio with respect to the number of executions, while the latter should be reset every time a test case is selected again.

Dependencies and similarities between test cases are often considered to make informed decisions during TCP. On the one hand, dependencies refers to pair of test cases that should be considered together (\textit{join execution})~\cite{Che11}, \textit{implementation dependencies} among them~\cite{Hai13}, or interrelated outcomes (\textit{verdict patterns}) expressed as association rules~\cite{Pra18}. On the other hand, test case similarity is defined as the distance between test cases based on a specific criterion, \eg \textit{code}~\cite{Noo17}, \textit{inputs}~\cite{Led12} or \textit{text} (comments, identifiers and literals)~\cite{Tho14}. The underlying idea is that choosing diverse test cases somehow guarantees that different functionalities are being tested. \textit{Code similarity} is an example of an attribute that can be extracted from more than one artefact (source code or traces). Both dependencies and similarities are often derived from a previous pairwise analysis of test cases, and thus require dealing with multiple values (arity 1:N or N:N). Although it can be viewed as costly, these attributes do not experience any change unless new test cases are incorporated to the test suite. In such a case, the pairwise values will grow incrementally, so that the greatest effort occurs when preparing data for the first implementation of the TCP technique.

\subsection{SUT information}\label{subsec:group2}

\begin{table*}[ht]
\caption{Classification of attributes based on the system under test.}
\label{tab:eval-sut-summary}
\begin{tabular}{lllllll}
\toprule
\multicolumn{2}{l}{\textbf{Source}} & \multicolumn{3}{l}{\textbf{Collection}} & \multicolumn{2}{l}{\textbf{Representation}} \\
\toprule
\emph{Entity} & \emph{Attribute} & \emph{Artefact} & \emph{Observation} & \emph{Session} & \emph{Type} & \emph{Arity} \\
\midrule
Class
& CBO               & SC       & Derived & Evolving      &Num.  & 1:N \\
& Class size        & SC       & Simple  & Evolving      &Int.  & 1:1 \\
& DIT               & SC       & Simple  & Evolving      &Int.  & 1:N \\
& Fault-proneness   & BC       & Derived & Evolving      &Num.  & 1:N \\
& Invocations       & BC       & Derived & Evolving      &Num.  & 1:N \\
& LCOM              & SC       & Derived & Evolving      &Num.  & 1:N \\
& NMI               & SC       & Simple  & Evolving      &Int.  & 1:1 \\
& NMO               & SC       & Simple  & Evolving      &Int.  & 1:1 \\
& NOC               & SC       & Simple  & Evolving      &Int.  & 1:N \\
& PIM               & SC       & Simple  & Evolving      &Num.  & 1:1 \\
& RFC               & SC       & Derived & Evolving      &Num.  & 1:N \\
& WAC               & SC       & Derived & Evolving      &Num.  & 1:1 \\
& WMC               & SC       & Derived & Evolving      &Num.  & 1:1 \\
\midrule
Component
& Instability       & Spec.     & Derived  & Evolving   &Num.  & 1:N \\
\midrule
Program
& Cohesion          & SC       & Derived  & Evolving    &Num.  & 1:1 \\
& Complexity        & SC       & Derived  & Evolving    &Num.  & 1:1 \\
& Frequency of use  & Spec.     & Simple  & Static      &Cat. & 1:1 \\
& Type of system    & Spec.     & Simple  & Static      &Cat. & 1:1 \\
& Version           & Spec.     & Simple  & Static      &Int. & 1:1 \\
\midrule
Inputs
&Data patterns      & BC        & Derived & Static      &Num.      & N:N \\
&Parameter values   & Spec.     & Simple  & Static      &Nom./Num.  & 1:1 \\ 
\midrule
Change
& Buggy change      & SC      & Derived  & Evolving   &Bin.  & 1:N \\
& Change intensity  & BC      & Derived  & Evolving   &Num.  & N:N \\
\bottomrule
\end{tabular}
\begin{tablenotes}
\item \footnotesize{Artefact: BC=Binary code, Spec=Specification, SC=Source code}
\item \footnotesize{Type: Bin=Binary, Cat=Categorical, Int=Integer, Nom=Nominal, Num=Numerical}
\end{tablenotes}
\end{table*}

Table~\ref{tab:eval-sut-summary} shows the list of information attributes of the \textit{SUT information} group, which serves to adapt the TCP technique to the particularities of the tested system. Here, the information entities refer to aspects regarding the invocation and evolution of the SUT at different levels of abstraction. SUT-related information for TCP is mostly obtained by parsing the source code to an abstract representation (70\%), \eg syntax tree or dependency graphs. A few attributes at the component or program levels rely on the specification, whereas binary code instrumentation is required for three attributes (\textit{change intensity}, \textit{invocations} and input \textit{data patterns}). From these observations, it can be concluded that the application of SUT-oriented TCP techniques is highly limited if source code is not available or it cannot be fully instrumented.

Among the code elements from which an attribute is computed, classes constitute the finest-grained entity with 13 metrics currently used in the literature. The six metrics defined by Chidamber and Kemerer~\cite{Chidamber94} have appeared in TCP studies~\cite{Do08,Mir08,Tou18,Sin19}. The \textit{class size}~\cite{Tou18} and an estimation of its \textit{fault-proneness}~\cite{Pat19} are less common attributes. In general, all these numerical metrics measure the complexity and level of dependency of classes, guiding the selection of those fault-prone modules that should be tested first~\cite{Do10,Tou18}. Similarly, component \textit{instability}, \ie a ratio between required and provided components has been used for TCP at a higher level of abstraction~\cite{Sil19}. 

The rest of code-oriented metrics refer to the whole program studied at different granularity levels. For instance, program \textit{cohesion} has been defined based on package, component, class and method-level analysis~\cite{Pan17}. One TCP technique incorporates the \textit{type} of system, with respect to its installation procedure, and the \textit{frequency of use}, linked to the importance of different parts of the system, in its decision process~\cite{Li13}. Both attributes are the only categorical attributes within this group, whose values are assigned by experts~\cite{Li13}. Similarly to the type of test case, additional values than those originally proposed by the authors could be defined. Given that classes, components and programs, \ie the main entities for SUT-related information, can be decomposed into smaller units, several attributes are derived from the analysis of such units and therefore have arity 1:N. All these attributes are denoted as evolving, meaning that their values do not depend on whether a test case is finally executed or not, but they are subject to SUT modifications.

Two other entities are not mapped to pieces of code, but they also provide SUT-related information. On the one hand, program inputs, expressed in form of either \textit{data patterns} or \textit{parameter values}, have been analysed to induce those values that make test cases fail~\cite{Wan12}. SUT \textit{inputs} have been studied for small programs with numeric or string parameters~\cite{Wan12,Rop19}. On the other hand, \textit{buggy change} estimations~\cite{Tan15} and the \textit{change intensity}~\cite{Mir07}, \ie semantic similarity between program versions, are derived observations that can be included in TCP models to guide the process towards recent SUT modifications. The former is predicted from 18 change metrics~\cite{Tan15}, whereas the latter compares two versions of the program to give a similarity score~\cite{Mir07}.

\subsection{Relational information}\label{subsec:group3}

\begin{table*}[ht]
\caption{Classification of attributes based on relational information.}
\label{tab:eval-relational-summary}
\scalebox{0.88}{
\begin{tabular}{lllllll}
\toprule
\multicolumn{2}{l}{\textbf{Source}} & \multicolumn{3}{l}{\textbf{Collection}} & \multicolumn{2}{l}{\textbf{Representation}} \\
\toprule
\emph{Entity} & \emph{Attribute} & \emph{Artefact} & \emph{Observation} & \emph{Session} & \emph{Type} & \emph{Arity} \\
\midrule
Test
&Additional coverage    &BC/SC/Tra.    &Derived &Dynamic  &Int./Num. &1:N \\
case
&Change coverage        &BC/SC/Tra.    &Derived &Evolving  &Bin./Num. &1:N \\
coverage
&Complexity coverage    &BC/Spec.       &Derived &Static   &Num.      &1:N \\
&Configuration coverage &Spec.          &Derived &Static   &Num.      &1:N \\
&Coverage distance      &BC/Spec.       &Derived &Static   &Num.      &N:N \\
&Coverage frequency     &BC             &Derived &Dynamic  &Int.      &1:N \\
&Coverage percentage    &BC/Spec.       &Derived &Static   &Num.      &1:N \\
&Coverage profile       &BC/Spec./Tra.  &Derived &Static   &Bin./Num. &1:N \\
&Database coverage      &BC/Spec.       &Derived &Static   &Bin.      &1:N \\
&Functional coverage    &BC/Spec./Tra.  &Derived &Static   &Bin./Num. &1:N \\
&GUI coverage           &Exec.          &Derived &Static   &Num.      &1:N \\
&Historical coverage    &BC/Spec.       &Derived &Dynamic  &Num.      &N:N \\
&Input coverage         &BC             &Derived &Static   &Int.      &1:N \\
&MC/DC coverage         &BC             &Derived &Static   &Bin.      &1:N \\
&User-defined coverage  &Spec.          &Simple  &Static   &Cat.      &1:1 \\
&Weighted coverage      &SC/Tra.        &Derived &Evolving &Num.      &1:N \\
\midrule
Invocation
&Active blocks          &BC/Tra.      &Derived  &Static   &Cat./Num.  &1:N  \\
&Change calls           &BC/SC        &Derived  &Evolving &Int.       &1:N  \\
&Failing calls          &Tra.         &Derived  &Dynamic  &Num.       &N:N  \\
&Method calls           &SC.          &Derived  &Static   &Num.       &1:N  \\
&Relevant statements    &Tra.         &Derived  &Static   &Num.       &1:N  \\
&Sequence calls         &Tra.         &Derived  &Static   &Nom.       &1:N  \\
\midrule
Program
&System functions       &Spec.      &Simple  &Static    &Num.       &1:N  \\
&Test inputs            &Spec.      &Simple  &Static    &Num.       &1:N  \\
&Test outputs           &Exec.      &Derived &Evolving  &Num.       &1:N  \\
&Usage patterns         &Tra.       &Derived &Evolving  &Num.       &N:N  \\
\midrule
Risk
&Component risk         &SC/Spec.   &Derived &Static    &Num.       &1:N  \\
&Risk coverage          &BC         &Derived &Static    &Num.       &1:N  \\
&Risk exposure          &Spec.      &Simple  &Static    &Bin./Int.  &1:N  \\
\midrule
Fault
&Fault age              &Exec./Spec.    &Derived &Evolving &Cat./Num.  &1:N \\
&Fault count            &Exec.          &Derived &Dynamic &Num.        &1:N \\
&Fault index            &BC/Spec        &Derived &Dynamic &Num.        &1:N \\
&Fault probability      &BC             &Derived &Static  &Num.        &1:N \\
&Fault severity         &Exec./Spec.    &Derived &Static  &Cat./Num.   &1:N \\
&Killed mutants         &Exec.          &Derived &Dyn./Sta.  &Num.     &1:N \\
\midrule
Change
&Change frequency       &SC/Spec.   &Derived &Evolving &Num.  &N:N  \\
&Changed methods        &BC/SC      &Derived &Evolving &Int.  &1:N  \\
&Failing issues         &Spec.      &Derived &Dynamic  &Num.  &N:N  \\
&Issue score            &Spec.      &Derived &Dynamic  &Num.  &N:N  \\
&Project changes        &BC         &Simple  &Evolving &Bin.  &1:N  \\
&Text score             &SC         &Derived &Evolving &Num.  &1:N  \\
\bottomrule
\end{tabular}
}
\begin{tablenotes}
\item \footnotesize{Artefact: BC=Binary code, Exec=Execution results, SC=Source code, Spec=Specification, Tra=Traces}
\item \footnotesize{Type: Bin=Binary, Cat=Categorical, Int=Integer, Num=Numerical}
\end{tablenotes}
\end{table*}

This group presents not only the largest list of information attributes, but also the broader range of entities, as shown in Table~\ref{tab:eval-relational-summary}. This fact confirms that many TCP methods need to establish connections between test cases and the SUT, but that these connections might come from very different places. In fact, it is the only category for which all types of artefacts appear, and several attributes can even be obtained from more than one artefact. Having alternative artefacts makes the attribute ---and therefore the corresponding TCP method--- far more flexible, allowing choosing depending on how costly the collection process is. The joint analysis of test and code elements also explains the prevalence of derived observations, whose arity tends to be 1:N (80\%) or N:N (17\%).

As expected, test case coverage is the most popular relational information asset for TCP in the literature~\cite{Elb00}, and stands out as the category for which more artefacts have been explored, from specifications to traces. Up to 16 different information attributes have been identified, which are expressed as a ratio of elements covered (numerical type) or as a list of values specifying whether each element is covered or not (binary type). Notice that almost all coverage attributes are derived and have arity 1:N, since they require identifying the elements , \ie statements, methods, branches, etc., exercised by each test case prior to the computation of the coverage value. \textit{Functional coverage} has been defined at multiple levels, the most common being statement, block and method. Other formulations take configuration~\cite{Qu07}, database entities~\cite{Ros17}, GUI steps~\cite{Ngu19}, or conditions, a.k.a \textit{MC/DC coverage}~\cite{Jon01}, as the elements to be covered. It is even possible to find a \textit{user-defined coverage} function~\cite{Tah16}, the only coverage attribute obtained from a single value (arity 1:1) and specified by an expert. Computing coverage only for the test cases not selected yet, \ie the ``additional approach''~\cite{Rot99} has been applied to functional, MC/DC and change coverage. Other attributes use coverage information to quantify the parts of the SUT that are more relevant to each test case. Some examples are \textit{coverage distance}~\cite{Jia12}, \ie distance of test cases based on their coverage, and \textit{coverage profile}~\cite{Leo03}, that retrieves the groups of statements and branches covered by the test case.

The analysis of calls from test code to system code constitutes another way to establish the connection between test cases and the functionality under test. Attributes under the invocation entity are derived from various values (arity 1:N or N:N) but, in contrast to coverage attributes, they do not access to SUT code because they only inspect how test cases make invocations. We found attributes that indicate when and how often a test case executes a code block (\textit{active blocks})~\cite{Ule18} or the \textit{sequence calls} executed by a test case~\cite{Rop19}, among others. It is also possible to prioritise test cases based on the number of \textit{method calls} as a surrogate of method coverage~\cite{Zha09-icsm} or to give more importance to certain statements identified by the tester (\textit{relevant statements})~\cite{Jef06}.

Next entity in Table~\ref{tab:eval-relational-summary} is the program, for which four attributes have been found. The \textit{inputs} that the test case passes to the system and the \textit{outputs} received are used to cluster similar test cases prior to prioritisation~\cite{Len13}. Also, the name and number of functions associated to system tests (\textit{system function})~\cite{Ala18} have been applied to TCP to estimate reliability of a safety-critical system, whereas the similarity between \textit{usage patterns}~\cite{And19} allows including the impact of faults on different users in the TCP process. Some program-related attributes (\textit{test outputs} and \textit{usage patterns}) are expected to evolve as the SUT does.

The last three entities, namely risk, fault and change, are more related to the project specification and evolution. Information coming from these entities is usually derived and present a multiple arity (1:N or N:N), since the attributes need to determine how the test case is affected by a set of risks, faults or changes. When risks are properly identified and documented, it would be possible to prioritise test cases based on its \textit{risk coverage}~\cite{Wan18} or to sort them according to the criticality of the components that each test covers (\textit{component risk})~\cite{Sil19}. Test cases can also be ordered with respect to their ability to detect risky faults (\textit{risk exposure})~\cite{Yoo11}. However, this attribute assumes that faults have been related to risks beforehand. Notice that risk estimations are static, \ie remain unchanged, as they are derived from the specification. Fault-related information is widely studied for TCP according to our literature analysis. Among others, fault \textit{age}, its \textit{probability} of occurrence and its \textit{severity} should be mentioned~\cite{Zha13,Lac16}. \textit{Fault age} and \textit{severity} might be associated to pre-established categories~\cite{Lac16} coming from specifications. The number (or ratio) of exposed faults (\textit{fault count})~\cite{Lac16} and \textit{mutants killed}~\cite{Rot99} are other indicators of the test case detection capability. An estimator of the fault-proneness of the functions covered by the test case has been developed too~\cite{Elb00}.

Focusing on change-based information, the content and priority of requests and issues have been studied as drivers of the TCP process. For instance, word similarity between test cases and changed files (\textit{text score}) or \textit{failing issues} is a mechanism to identify relevant test cases~\cite{Bus16}. Also, test cases affected by changes in the methods they cover (\textit{changed methods})~\cite{Hua09} or linked to added/removed/modified project artefacts (\textit{project changes})~\cite{Ule18} between consecutive program versions can be identified and prioritised. Attributes related to faults and changes have an evolving or dynamic nature depending on whether their formulation is subject to the SUT lifetime (\textit{project changes} and \textit{fault age}) or it is coupled to the session (\eg \textit{failing issues} and \textit{fault count}), respectively.

\subsection{Taxonomy evaluation}\label{subsec:eval}

Since the \acronym taxonomy was created bottom-up, based on analysing the literature, there is a risk that it might not generalise to papers\slash methods outside of this group. While we do foresee and expect future updates to the taxonomy as the area evolves, \acronym should provide now a framework consistent with the current development of the area and therefore these updates are expected to be gradual and sustained over time. To assess the generalisability of the proposed taxonomy, here we apply it to a set of studies not used during the development phase, \ie the TCP studies published in 2020. Again, we run the three search strings described in Section~\ref{sec:methodology} for 2020, respectively retrieving 13, 3 and 51 papers.\footnote{Source: Scopus (December 4, 2020).} From the total of 61 distinct papers, we select 9 articles following the same methodology explained in Section~\ref{sec:methodology}.

\begin{table*}[ht]
\caption{Evaluation of the \acronym taxonomy with TCP studies published in 2020. The column marked `S' indicates the severity of the taxonomy update\slash adaptation prompted by a paper (1 indicating no update, 2 some adaptation).}
\label{tab:taxonomy-evaluation}
\begin{tabular}{lllllp{2cm}}
\toprule
\textbf{Reference} & \textbf{Group} & \textbf{Attribute} & \textbf{S} & \textbf{Comments} \\ 
\toprule
Bertolino                   & Test. Inf. & Execution time & 1 & \\
et al.~\cite{Ber20}         & Rel. Inf. & Fault probability & 2 & - Estimated from 50 metrics\\
& &  &  & - Extracted from different sources\\
\midrule
Chi et al.~\cite{Chi20}     & Rel. Inf.  & Coverage frequency    & 1 & \\
                            & Rel. Inf.  & Method calls          & 2 & - Use of weights \\
                            &                   &                       &   & - Extracted from traces \\
\midrule
Huang et al.~\cite{Hua20}   & Rel. Inf.  & Configuration coverage & 1 & \\
\midrule
Huang et al.~\cite{Hua20jss}&Rel. Inf. & Additional coverage & 1 & - For code unit combinations\\
\midrule
Lam et al.~\cite{Lam20}     & Test. Inf. & Previous priority & 1 & \\
                            & Test. Inf. & Joint execution & 2 & - Representing order dependencies\\
                            &  &  &  & - Extracted from execution\\
                            & Rel. Inf. & Additional coverage & 1 & \\
                            & Rel. Inf. & Functional coverage & 1 & \\
\midrule
Lu et al.~\cite{Lu20}       & Rel. Inf. & Additional coverage & 1 & \\
\midrule
Mahdieh                     & SUT inf.          & Fault-proneness       & 2 & - More metrics for estimation \\
et al.~\cite{Mah20}         &                   &                       &   & - Extracted from source code \\
                            & Rel. Inf.  &  Additional coverage  & 1 & \\
                            & Rel. Inf.  &  Functional coverage  & 1 & \\
\midrule
Wang et al.~\cite{Wan20}    & Rel. Inf.  & Coverage distance     & 1 & - New distance function\\
\midrule
Xiao et al.~\cite{Xia20}    & Test. Inf.      & Effectiveness         & 1 &\\
                            & Test. Inf.      & Execution time        & 1 &\\
                            & Test. Inf.      & Historical verdicts   & 1 &\\
\bottomrule
\end{tabular}
\end{table*}

We considered that three situations might occur when classifying the attributes appearing in these papers using the \acronym taxonomy. Firstly, the attribute fits the definition and characteristics specified by the taxonomy, so no update is needed (scenario \#1). Secondly, the attribute appears in the taxonomy but it is handled slightly differently in the paper (scenario \#2). In this case, some categories might need to be adapted to capture the new proposed ideas. Lastly, if the attribute has not been defined in the taxonomy, it has to be fully characterised and added to the taxonomy (scenario \#3). In terms of generalisability of the taxonomy, scenario \#1 would be the most favourable. Under scenario \#2, it could be necessary to extend some attribute definitions or assign additional values to the affected categories in order to better reflect the variety of approaches. Table~\ref{tab:taxonomy-evaluation} lists the attributes appearing in each of the five TCP papers we found published during 2020, including the corresponding group and the evaluation scenario (S) we judge them to lead to. The last column of the table enumerates characteristics that might imply extensions in some attribute categories. Next, we briefly describe each study, explaining how these observations affect the taxonomy:

\begin{itemize}
    \item Bertolino et al.~\cite{Ber20} compare the performance of ten ML algorithms to estimate the fault probability of classes, combining this information with the execution time to prioritise test cases. Both attributes appear in the taxonomy. Fault probability is estimated from 50 metrics classified into program size, cyclomatic complexity, object-oriented metrics and test history. Due to the variety of metrics, new sources---\eg source code and execution results---are required to compute the information attribute (scenario~\#2). In addition, the information attribute could be considered as evolving since the ML algorithms are trained after each commit.
    
    \item Chi et al.~\cite{Chi20} use \textit{method calls} as primary attribute for TCP. Calls are extracted from traces and weighted depending on whether they represent code-to-code, test-to-code or test-to-test invocations. \textit{Coverage frequency} is applied as secondary criterion in case of tie, so that elements less frequently covered obtain higher priority. \textit{Method calls} appears in the taxonomy as a relational attribute extracted from source code instead of traces (see Table~\ref{tab:eval-relational-summary}), meaning that an additional source can be considered (scenario~\#2). Also, the novel use of weights implies that this attribute could be grouped now under testing information (close to the attribute \textit{implementation dependencies}) and SUT information (equivalent to \textit{invocations}). This is the first case of an attribute being classified in multiple groups. On the other hand, \textit{coverage frequency} coincides with the definition provided by the taxonomy (scenario~\#1).
    
    \item Huang et al.~\cite{Hua20} define a TCP technique for combinatorial testing. The method compares different levels of input parameter coverage to prioritise test cases. The approach matches with the \textit{configuration coverage} defined in the \acronym taxonomy (scenario~\#1).
    
    \item Huang et al.~\cite{Hua20jss} propose a new \textit{additional coverage} measure based on the number of code unit combinations that are covered by a test case, among those not covered yet. The approach is evaluated at branch, method and statement level. This work fits into scenario~\#1, since the definition of the information attribute is not subject to the specific coverage formulation, \ie it could be added to the list of available approaches (functional, change and MC/DC coverage).
    
    \item Lam et al.~\cite{Lam20} combine testing and SUT-oriented information attributes, all of them already defined in the taxonomy. The attribute \textit{joint execution} has been identified as scenario~\#2 since the type of dependency between test cases refers to the order in which test cases should be executed, which is determined by automatic tools. In the original definition, the dependencies were manually specified by tester and could express any kind of dependency.
    
    \item Lu et al.~\cite{Lu20} proposes a new search algorithm that is guided by additional coverage, so scenario~\#1 is applied here.
    
    \item The method by Mahdieh et al.~\cite{Mah20} gives more priority to those test cases that cover fault-prone classes. The fault probability serves to define a modified coverage function, an idea applied to the additional coverage approach too. The estimation is based on 104 metrics: 52 source code metrics, 8 clone metrics, 42 metrics for coding rule violation and 2 for Git history. The number and variability of metrics for \textit{fault-proneness} estimation is higher compared to the study from which this attribute was identified during the taxonomy development~\cite{Pat19}. However, this level of detail is not included in the attribute characterisation, so only source code has to be added to the list of sources according to scenario~\#2.
    
    \item The method by Wang et al.~\cite{Wan20} is based on pair-wise test case similarity. In particular, they define a novel coverage distance function based on the XOR (exclusive OR) operator. Scenario~\#1 is applied since the specific distance function is not represented in the taxonomy.
    
    \item Xiao et al.~\cite{Xia20} apply deep learning to predict the fault detection capability of test cases. The approach is based on test information only, using the same attributes and data sets that other ML-based studies considered for building the \acronym taxonomy~\cite{Spi17,Wu19} (scenario~\#1).
\end{itemize}

This initial evaluation exemplifies the capacity of the taxonomy to describe the information attributes used by TCP methods. On a positive note, none of the evaluated papers led to scenario~\#3. The fact that a general attribute definition was derived from papers adopting similar concepts reduces the risk of leaving aside variations yet to be explored. After evaluating two TCP studies~\cite{Chi20,Mah20}, we have identified different causes that would imply changes in two categories (scenario \#2), namely group and source. On the one hand, we have observed that a same attribute (\textit{method calls}) can also be classified as SUT-information (if invocations happen in the SUT) or testing information (when invocations are extracted from the test cases).
Our view is that this phenomenon will not occur often and if so, the attribute can be associated to multiple groups. On the other hand, multiple values for the source category were conceived from the beginning to deal with the diverse situations reported in TCP studies. Together with the source, we anticipate that the type of variable could be another category experiencing this type of extension, since multiple values appeared during the taxonomy development too. However, type diversity mostly responds to authors' decisions on how the value is best stored. Finally, the categorical values assigned to certain attributes, especially those describing characteristics of test cases like its type or status, are expected to change or be extended to better reflect the reality of different companies. 

As part of our interviews with software QA professionals, we ask them to provide feedback about the \acronym taxonomy and its benefits. The first company considered that artificial intelligence can be of great help in testing automation, and the presented ideas have value in making such possibility more tangible. The second company highlighted the importance of having well-defined metrics that, as part of an automation tool and combined with context information about the SUT, could greatly support the testing process. Finally, the third company expressed that the \acronym taxonomy might be useful to define the testing strategy, as it provides an overview of available options, but that such strategy should include organisational factors too.

%--------------------------------
% 6 - DISCUSSION (RQ2)
%--------------------------------
\section{Analysis of attributes used in industrial studies}
\label{sec:rq2_analysis}

In response to RQ2, this section analyses which information attributes have been used in industrial studies. Our aim is to discover if certain attributes are preferred or are considered in combination more frequently than others, as well as to identify current limitations with respect to the availability of information for TCP. To understand these factors, we carry out a more in-depth analysis of industry-oriented TCP studies. From the list of references considered to build and evaluate the taxonomy, 23 works (21\%) report industrial case studies or validate their proposals with data sets from industrial projects. Although these studies might not reflect all the requirements of TCP in practice, they usually describe a case study where the application context and decisions made are detailed. From that information, we can better understand why some information attributes were selected over others, or which are the context factors limiting the use of certain attributes. Also, the interviews with companies were conducted with two aims: 1) to contrast the findings regarding the relevance of information attributes, and 2) to complement the analysis about the observed limitations while reflecting on their testing experiences.

\subsection{Information attributes appearing in industrial studies}
\label{subsec:rq2_attributes}

The \acronym taxonomy defines three groups of attributes according to the type of information used: testing information (27 attributes), SUT information (23) and relational information (41). From the total set of 91 attributes, only 39 appear in TCP methods applied to industry projects. The distribution among groups is as follows: testing information (67\%), SUT information (13\%) and relational information (44\%). Table~\ref{tab:attributes_ind} provides the number of attributes belonging to each group and entity, as well as the percentage of them appearing in at least one industrial study. The last column lists the references using information attributes from the corresponding category (see the additional material for a detailed list of the attributes used by each reference).

\begin{table*}[ht]
\caption{Percentage of attributes considered by TCP methods evaluated in industry contexts. Asterisk indicates evaluation with already prepared datasets only.}
\label{tab:attributes_ind}
\begin{tabular}{llcccp{2.5cm}}
\toprule
\textbf{Group} & \textbf{Entity} & \textbf{No. Attr.} & \textbf{\% Attr.} &\textbf{No. Papers}&\textbf{References}\\
\toprule
Testing & Test case (report) & 2 & 100\% & 5 & \cite{Ala18},  \cite{Ali19}*, \cite{Bus16}, \cite{Spi17}*, \cite{Xia20}*\\
information & Test case (history) & 6 & 83\% & 12 & \cite{Bus16,Eng11,Lia18,Mar13,Mar15,Pra18,Pra19}, \cite{Spi17}*, \cite{Ule18,Wan16}, \cite{Wu19}*, \cite{Xia20}*\\
& Test case (execution) & 5 & 80\% & 10 & \cite{Hua09,Mar13,Mar15,Pra19}, \cite{Spi17}*, \cite{Sri02,Tah16,Ule18,Wan16}, \cite{Xia20}*\\
& Test case (dependency) & 3 & 67\% & 3 & \cite{Hai13,Pra18,Pra19}\\
& Test case (property) & 8 & 62\% & 4 & \cite{Bus16,Eng11,Tah16,Wan16}\\
& Test case (similarity) & 3 & 0\% & &\\
& \textit{Total} & 27 & 67\% & 18& \\
\midrule
SUT & Program & 5 & 40\% & 1& \cite{Li13}\\
information & Class & 13 & 8\% & 1 & \cite{Car11} \\ 
& Change & 2 & 0\% & & \\
& Component & 1 & 0\% & &\\
& Inputs & 2 & 0\% & & \\
& \textit{Total} & 23 & 13\% & 2 &\\
\midrule
Relational 
& Change & 6 & 67\% & 5 & \cite{Ali19}*, \cite{Bus16,Hua09,Ule18}\\
information
& Fault & 6 & 50\% & 4 & \cite{Car11,Hai13,Lac16,Mar15}\\
& Program & 4 & 50\% & 2 & \cite{Ala18,And19}\\
& Invocation & 6 & 33\% & 2 & \cite{Hua09,Ule18}\\
& Risk & 3 & 33\% & 1& \cite{Li13}\\
& Test case (coverage) & 16 & 31\% & 10 & \cite{Ali19}*, \cite{Bus16,Car11,DiN13,DiN15,Hua09,Lac16,Mar15,Sri02,Tah16}\\
& \textit{Total} & 41 & 44\% & 16 &\\
\bottomrule
\end{tabular}
\end{table*}

Within the testing information group, test case report is the only entity from which all attributes (\textit{effectiveness} and \textit{failure frequency}) have been employed in industry-oriented studies. Most of the attributes related to the history (83\%) and execution (80\%) of test cases seem valuable for industrial TCP and supported by a wider range of methods looking at the number of references. In particular, the \textit{historical effectiveness} and the \textit{execution time} are the attributes more frequently appearing in each category. Test case dependencies (67\%) have been used in a few studies that trace dependencies from their specification~\cite{Hai13} or analyse coincidences of verdicts~\cite{Pra18,Pra19}. The cost associated to similarity-based attributes might be one of the reason of their lack of application in industry. They all have N:N arity, requiring source code analysis and handling textual data. Finally, a considerable number of test case properties (62\%) have been included in TCP methods for industry, but none of them appear in more than one reference.

SUT-related information presents the lowest percentage of industrial application, with only two studies and 13\% of attributes. Difficulties in accessing the SUT code, as pointed out by some authors~\cite{Eng11,Henard16}, might be the reason of its low applicability. A closer look at the two studies that use class (8\%) and program (40\%) attributes seems to confirm this hypothesis. On the one hand, the method that incorporates class information (\textit{invocations}, \textit{functional coverage} and \textit{coverage distance}) needs code instrumentation, but the process was previously implemented by the company~\cite{Car11}. On the other hand, the program characteristics (\textit{type of system} and \textit{frequency of use}) are elicited from experts' judgement~\cite{Li13}.

As for relational attributes, change-based attributes present the highest percentage of application in industry (67\%), but we did not found any attribute used in more than study. \textit{Text score} and \textit{change frequency} are attributes applied in industry that compute similarities. The presence of both attributes contrasts with our previous finding about the absence of industrial application of attributes based on test case similarity (see testing information in Table~\ref{tab:attributes_ind}). Our thought is that similarity between test cases with respect to changes limits the scope of information to be managed, making it more practical. Only a few coverage functions (31\%) have been applied in industry compared to the wide range of formulations found in the literature. \textit{Functional coverage} and \textit{change coverage} appear more frequently, sometimes under the additional approach. As for fault-based attributes (50\%), the three attributes evaluated in industry are \textit{age}, \textit{severity} and \textit{fault count}, \ie number of failures detected by each test case. In all cases, the values were previously tracked or estimated by the company~\cite{Hai13,Lac16,Mar15}. Percentages of application for the remaining entities (invocation, risk and program) are below 50\%. Besides, these entities present less attributes and few studies report their use, from which we infer that their presence responds to the particular interests of the industrial partner.

We have contrasted these findings with the information gathered from the interviews with industry professionals. It should be noted that these companies perform RT, but their strategies are not fully automated and some aspects are constrained by the specific project, the level of testing and the dependencies with third-parties. Despite this, all participants were open to discuss the relevance of each group of attributes when making decisions, even if they do not currently apply them all for the above reasons. Focusing on testing information, all the entities are viewed as applicable, with more relevance assigned to test case dependencies, test case history and execution, and properties like age, type of test and previous priority. As for SUT information, complexity was highlighted as an important code metric driving testing decisions by two of the companies. Change-related attributes, which might belong to the SUT or the relational information, were mentioned too. In general, the traceability between test cases and code is seen as necessary for planning testing strategies, guided by risk analysis, fault information and coverage. Compiling all their answers, we observe a high alignment with respect to the literature analysis summarised in Table~\ref{tab:attributes_ind}, although relational information was deemed as more relevant than testing information by the company practitioners. The reasons are the possibility of access to the code and the lack of a fully automated process from which test results can be stored and exploited. 

\subsection{Exploring the combination of information attributes}\label{subsec:rq2_combination}

This section analyses which information attributes are combined to build TCP in industry. Mostly, industrial studies rely on a single group of attributes, either testing information (36\%) or relational information (16\%), although combining attributes from both groups is relatively frequent (44\%). Authors from the selected papers argue that considering attributes of different nature was essential to cover the many critical factors that affect software testing~\cite{Li13}, as well as to increase testers' confidence on the test suite analysis~\cite{Eng11}. Indeed, the combination of two or three information sources has been reported as a common practice in the context of TCP methods for CI environments~\cite{PradoLima20}.\footnote{Their classification of information sources has only nine categories and one level, so they cannot be directly mapped to our groups and entities.} Similarly, our analysis reveals that prioritisation methods for industry employ between one and five attributes, suggesting that industrial TCP is typically focused on a reduced number of information attributes.

\begin{figure}[ht]
  \centering
  \includegraphics[width=\linewidth]{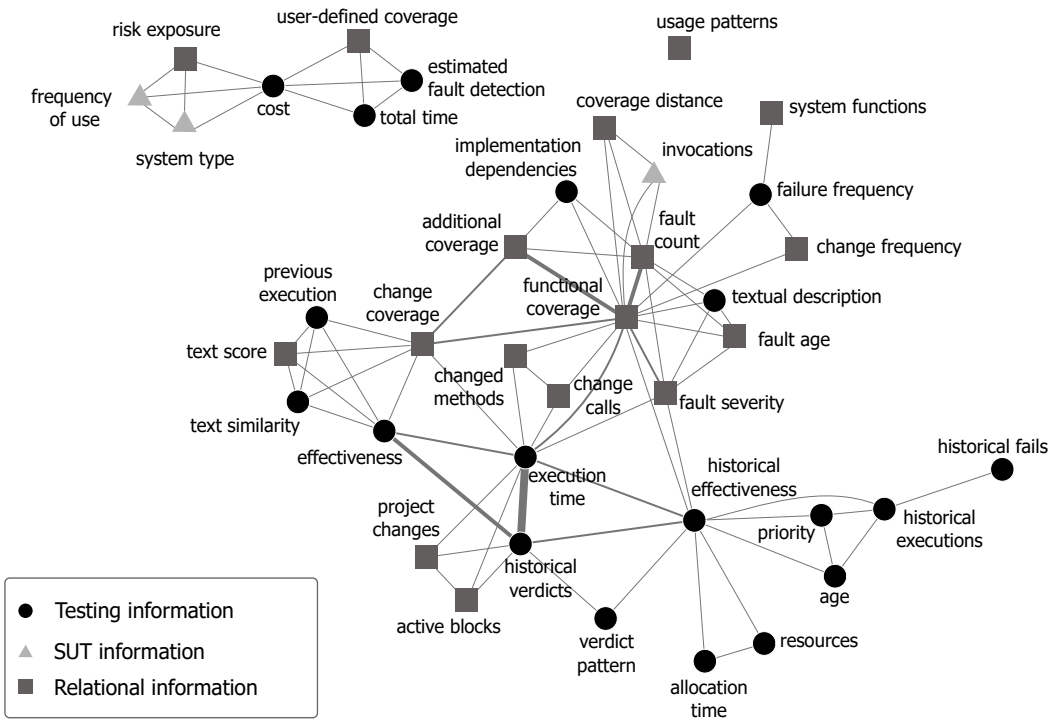}
  \caption{Attributes that have considered together in industry TCP studies.}
  \Description{A graph showing the attributes that have been considered in industry studies, and the connections between them.}
  \label{fig:graph-att-ind}
\end{figure}

Knowing which specific attributes have been combined allows practitioners to position each TCP method in relation to the information they usually manage for prioritisation. With this aim, Figure~\ref{fig:graph-att-ind} shows a graph establishing the connections between pairs of attributes, where the node shape and colour correspond to the attribute group, and the edge width represents the frequency of appearance (from 1 to 4 studies). Only one method, which analyses \textit{usage patterns}, does not combine attributes. \textit{Functional coverage}, \textit{execution time} and \textit{historical effectiveness} are the attributes for which more combinations have been explored. Looking at their edges, \textit{historical effectiveness} is more often combined with other history-based attributes (\textit{historical executions}, \textit{historical verdicts} and \textit{verdict patterns}) and test case properties (\textit{age} and \textit{priority}), 
whereas \textit{execution time} has been studied together with three change attributes (\textit{change coverage}, \textit{changed methods} and \textit{change calls}). \textit{Functional coverage} is also compatible with the same three change attributes plus the \textit{change frequency}, and is often analysed with fault properties (\textit{count}, \textit{frequency}, \textit{age} and \textit{severity}) too. It is interesting to note that some TCP works apply or compare several coverage attributes, thus giving more flexibility to adapt the TCP method to the preferred formulation.

\subsection{Observed limitations of TCP in industry}
\label{subsec:rq2_limitations}

In this section we summarise and discuss some limitations for the industrial application of TCP methods from the information perspective. More specifically, we have collected and classified the problems that guided the design of TCP methods in industry-oriented studies. Table~\ref{tab:ind_problems} presents the identified problems, as well as the proposed solutions related to information extraction and management. The table includes the references to the corresponding papers, where the asterisk indicates papers using already available industrial datasets. This implies that the related problems inspired the design of the TCP method, but they were not really addressed in practice. We also extract characteristics of the SUT that might influence TCP, such as domain, size and language, although we only refer to those cases relevant for our discussion.\footnote{Detailed information can be found in our GitHub repository (see additional material in page~\pageref{sec:additional}).}

\begin{table*}[ht]
\caption{Problems related to the selection of information attributes for TCP encountered in industry-oriented studies. Asterisk indicates evaluation with already prepared datasets only.}
\label{tab:ind_problems}
\begin{tabular}{lp{4cm}p{4.25cm}p{2.75cm}}
\toprule
\textbf{Category} & \textbf{Problem} & \textbf{Solution} & \textbf{References}\\
\toprule
Code        & Code analysis is expensive    & Focus on changes only & \cite{DiN13, DiN15, Hua09, Sri02}\\
analysis    & Coverage analysis is costly & Coverage at high level & \cite{DiN13, DiN15}\\
            & Exact metric computation is not possible & Estimated or approximated values & \cite{Bus16, Car11, Eng11, Hai13, Li13, Tah16}\\
            & No access to any form of code & Rely on historical results and test case properties & \cite{Ala18, Mar13, Mar15}, \cite{Xia20}* \\
            & No traceability between test cases and code & Rely on historical results and test case properties & \cite{Eng11}, \cite{Spi17}*\\
            & Source code is not available  & Work at binary level  & \cite{Bus16, Hua09, Sri02}\\
\midrule
CI      & Changes in the test suite & Adaptive TCP to allow adding and removing test cases & \cite{Ali19}*, \cite{Spi17}*, \cite{Wu19}*, \cite{Xia20}*\\
        & Fast change frequency & Asynchronous preprocessing & \cite{Bus16}\\
        & Time constraints & Evaluation under time limits & \cite{Car11}, \cite{Wu19}* \\
        &                  & Execution time as attribute & \cite{Mar13,Mar15}, \cite{Spi17}*\\
\midrule
Testing & Acceptance testing level & Definition of project-oriented & \cite{Li13}\\
process & & attributes & \\
        & Hardware preparation & Attributes for allocation time &  \cite{Tah16,Wan16}\\
        & Lack of business perspective  & Choose attributes aligned to business factors & \cite{Mar15,Tah16}\\
        & System testing level & Rely on information used in manual testing & \cite{Lac16}\\
        & Test case dependencies & Include dependency analysis & \cite{Hai13, Pra18, Pra19}\\
\midrule
SUT     & Faults affect different users & Definition of usage patterns as & \cite{And19}\\
charact.& or functionalities &  attributes & \\
        & Heterogeneity of artefacts and languages& Combination of attributes and ad-hoc preprocessing& \cite{Bus16} \\
        & Real-time conditions and specific standards & Information at runtime during testing & \cite{Ule18}\\        
\bottomrule
\end{tabular}
\end{table*}

A first problem category is related to the efforts devoted to code analysis~\cite{DiN15,Ule18}, especially if the procedures required to collect and process the information are not implemented by the company yet. Alternative solutions have appeared to mitigate the cost of code analysis, such as analysing only those parts of the SUT that have changed or computing coverage at the block level. Another recurrent approach is the use of estimated or approximate values, which can be obtained from testers~\cite{Li13}. Several authors explain that they opt for information attributes related to test case historical results and other properties (execution time, age, etc.) because access to the SUT code is not available in any form~\cite{Ala18,Mar13}, or the company does not trace the relationship between test cases and code artefacts~\cite{Eng11}. For those cases in which the company supports some kind of code instrumentation, the analysis of binary artefacts becomes a lighter process~\cite{Sri02}. Under this problem category, one important aspect is whether the information attributes remain static or have a dynamic or evolving nature. In the latter case, the cost-benefit of implementing a specific TCP method will depend on how frequently the information is expected to change and how often TCP is executed~\cite{Kim00,Minhas20}.

Most of the recent industrial studies address TCP under CI practices, which imposes additional constraints. Changes in the test suite are expected to happen more often, meaning that testing and relational information could become obsolete after adding, deleting or modifying test cases~\cite{Zha09-icsm}. Also, the relevance of each test case vary over time when testing is scheduled after every change~\cite{Xia20}, which led to the design of an adaptive ML-based method. Although these studies mention these factors as part of their motivation, they have not been evaluated in actual industrial settings. In contrast, time constraints are frequently considered when designing or evaluating TCP for CI environments. Two usual approaches are: 1) to analyse the TCP method effectiveness by defining different time limits~\cite{Car11}, and 2) to include test case execution time as an attribute for TCP~\cite{Mar13}, trying to maximise the number of test cases to be executed in each cycle.

The level of testing and the environmental settings also affect the definition of information attributes used for TCP.
Under controlled environments, test cases are unitary, independent and do not need special considerations to be executed. These assumptions do not hold in industrial case studies, as reflected by the appearance of attributes for inclusion in the taxonomy that were specifically conceived to introduce practicalities. Industry-oriented studies addressing acceptance and system testing include information attributes to represent project-specific information~\cite{Li13} and make use of the same information available in manual testing~\cite{Lac16}, respectively. Also, a business perspective can be introduced to decide which information attributes apply, \eg fault detection capability and test case execution are viewed as relevant for the customer experience~\cite{Tah16}. Two studies refer to the need of taking hardware preparation into account when designing their TCP methods. More specifically, new attributes for allocation time~\cite{Tah16} and resource utilisation~\cite{Wan16} were defined in their proposals, and also included in the taxonomy. Industrial collaborations have also highlighted the need of test case dependency analysis, using their results as information attributes~\cite{Hai13,Pra19}.

The characteristics of the SUT impose some adaptations of TCP and its input information attributes. For instance, a web application with a wide catalogue of functionalities and types of users has benefited from the analysis of usage patterns to locate fault-prone parts~\cite{And19}. This work was responsible for adding usage pattern as attribute in the taxonomy, as it was not considered in any non-industrial paper before. Another case study mentions that the heterogeneity of the artefacts comprising the SUT, which is written in multiple languages, complicates the extraction of usual information for TCP~\cite{Bus16}. The authors developed ad-hoc methods to process and combine the information from the available sources. Existing TCP methods also become inapplicable to test embedded systems under real-time conditions and strict control standards, which imposed runtime analysis of hardware behaviour. Overall, the characteristics of the SUT in terms of application domains (automotive, video conferencing, manufacturing, CRM), languages (C, C++, Java) and size (reaching up to 3M LOC) greatly differ from the small programs, \eg benchmarks from the SIR repository,\footnote{\url{https://sir.csc.ncsu.edu/portal/index.php} (Accessed January 14, 2022)} or open source projects, \eg those collected in the Defects4J repository,\footnote{\url{https://github.com/rjust/defects4j} (Accessed January 14, 2022)} often used for TCP research evaluation. Most of the TCP methods evaluated with these repositories might not scale well to industrial SUT, or would require considerable adaptations. Having a complex SUT also implies that its test suite will greatly differ from those available in public repositories. The industrial case studies analysed here report test suites with thousands of test cases~\cite{Ali19,Bus16,Sri02}, that takes several days or weeks to run~\cite{Ala18,Hua09,Li13}, and that do not always correspond to unit testing~\cite{Ala18,Ule18,Tah16}. In this sense, the cost of obtaining attributes with arity N:N will grow exponentially with the number of test cases, and its applicability will be ultimately conditioned by the involved artefacts and the efficiency of the extraction process.

Finally, we shared the list of problems with the software QA professionals during the interviews, asking them whether they experience the same or other problems in their daily work. Problems related to code analysis were deemed as representative for one company, since they have to deal with code from external teams that makes automation quite complex. The traceability cost was particularly highlighted by other participant, but it was not an impediment in the other (smaller) company. All these companies follow CI practices, with different opinions on their associated problems. The company adopting CI more recently acknowledges that maintenance and communication problems arise under CI, whereas the largest company mentions that the size of the test suite and the fast changes are challenging issues in this context. In terms of testing process, we got different views due to the variety of target areas in each of the three companies, as well as others in which the participants worked previously. The existence of different hardware environments was mentioned by two companies, resulting in both high preparation cost and implications in the test suite maintenance in one of them. Regarding the SUT, the interviews with the company practitioners confirm that the co-existence of code and other artefacts, as well as the use of multiple programming languages, complicates the testing process as mentioned above. Only one company has to adhere to specific standards and regulations due to their target domain. Additional problems are mostly related to communication and dependencies with third-party teams, maintenance costs, and the control imposed from the business perspective. From all these comments, we can conclude that several problems observed in the literature still persist in the industry, specially those related to the particularities of the testing process and SUT characteristics. In contrast, the availability of tools for testing automation and quality assurance seems to be mitigating past problems regarding code analysis and CI adoption. As pointed out by some of the QA professionals, it would be desirable to integrate TCP methods within the testing automation tools they are familiar with.

%--------------------------------
% 7 - DISCUSSION ML (RQ3)
%--------------------------------
\section{Analysis of attributes used in ML-based TCP}
\label{sec:rq3_analysis}

With the increasing interest in applying ML to TCP, it is important to analyse the information currently used for learning and whether it is aligned with the insights gathered from the analysis of industry-oriented publications. In this section, we first introduce the different learning tasks that have been incorporated to drive or support TCP. Then, we focus on the information attributes appearing in ML-based TCP studies, with special emphasis on those evaluated with industrial data. Based on this analysis, we discuss additional aspects that should be considered when adapting such attributes to become data features suitable for learning. 

To support the analysis, Table~\ref{tab:attributes_ml} summarises the percentage of attributes belonging to each entity that appears in ML-based TCP studies. The list of references is classified by type of learning approach, namely supervised, unsupervised, semi-supervised, reinforcement, deep, and multi-approaches (see Section~\ref{subsec:ml-testing}). For each one, references are further divided into industrial and non-industrial evaluation (mostly on open source projects). In total, we found 35 ML-based TCP studies, 10 of which report an industrial case study or an experimental validation with industrial data sets. In the context of ML, 53 information attributes from the taxonomy have appeared with the following distribution: 16 testing attributes (59\%), 14 SUT-oriented attributes (61\%) and 23 classified as relational information (56\%). 

\subsection{Learning approaches for TCP}\label{subsec:rq3_algorithms}

Although the scope of our analysis is the information collected for learning and not the specific ML algorithms, the learning approach can influence the type of attributes required by the algorithm. Therefore, we firstly identify the learning purposes of the TCP methods applying ML, which allows us to analyse if some attributes are more likely to appear under a particular learning approach. Next, we summarise the TCP applications of the learning paradigms\footnote{Details on the specific algorithms applied can be found in the GitHub repository available as additional material.} presented in Section~\ref{sec:background}:

\begin{itemize}
   
  \item \emph{Supervised learning}. A classification algorithm can be responsible of producing the rank of test cases~\cite{Bus16,Lac16,Jah19}, meaning that past priorities should be known to be used as labels. Learning a function to prioritise test cases based on pairwise comparisons requested to an user has been explored too~\cite{Ton06}. Another classification task is predicting whether a test case will fail or not~\cite{Noo17,Pal18}, so a prioritisation step is needed thereafter. Classification also serves to extract knowledge from the SUT that become the input for TCP techniques. Some examples are the identification of fault-prone classes~\cite{Ber20,Mah20,Tou18} and defective modules~\cite{Sin19}, the prediction of whether a change will cause test cases to fail~\cite{Tan15}, and the estimation of the execution probability of a part of the SUT~\cite{Wan12}.
   
  \item \emph{Unsupervised learning}. Clustering, especially hierarchical methods, acts as a selection mechanism prior to prioritisation. Once similar test cases are grouped, only those belonging to a particular cluster are executed~\cite{Pan13,Ros17}. Another possibility is defining intra- and inter-cluster procedures to order test cases~\cite{Ali19,Car11,Che18b,Fan14,Fu17,Kan17,Leo03,Yoo09,Zha15}. Rule mining has been applied to complement a TCP technique~\cite{Pra18,Pra19} with the purpose of extracting patterns from test case executions. Topic modelling has served to analyse test case descriptions, giving more priority to test cases related to specific topics~\cite{Tho14}.

  \item \emph{Semi-supervised learning}. This approach has been applied to group test cases detecting the same fault, letting the user to specify constraints about test cases that should be considered together~\cite{Che11}.

  \item \emph{Reinforcement learning}. All methods are oriented towards predicting the outcome of test cases. However, the reward function has been implemented with different techniques: neural networks~\cite{Spi17}, random forest~\cite{Ber20}, founded on fixed rules defined a priori~\cite{Ngu19} and using time windows~\cite{Wu19}.
   
  \item \emph{Deep learning}. The only study applying DL formulates the test case outcome prediction problem using temporal series~\cite{Xia20}. A recurrent neural network performs the classification of test cases, continuously learning from test case results that become available.
    
  \item \emph{Multi-approaches}. The outcomes of a clustering process are employed to label test cases. Depending on the sort of clusters and whether the labelling process is done automatically or manually, supervised learning~\cite{Bha18,Len13}, semi-supervised learning~\cite{Rop19} or active learning~\cite{Ala18} is executed thereafter.
\end{itemize}

The identified learning tasks require different groups of attributes as inputs, as shown in Table~\ref{tab:attributes_ml}. The greater number of supervised tasks is reflected in a wider range of attributes, covering almost all groups of attributes and entities. For the test case outcome prediction problem, a combination of testing and relational information is preferred~\cite{Bus16,Noo17,Pal18}. In contrast, predicting whether a component of the SUT will fail or not only requires SUT-related attributes, such as change and class metrics~\cite{Sin19,Ber20,Mah20}. Focusing on unsupervised approaches, clustering can analyse the similarities between test cases based on how they relate to different parts of the SUT (\ie relational information). Clustering methods usually measure coverage distance based on the coverage profiles built for each test case, \ie an array indicating whether a code construct is covered by the test case or not~\cite{Yoo09}. The joint execution of test cases, an attribute from the testing information group, represents the type of dependency to be discovered via rule mining~\cite{Pra18,Pra19}. Similarly, topic modelling only requires testing information, as it analyses textual specifications of test cases~\cite{Tho14}.

The rest of learning paradigms have been less explored for TCP, which is reflected in less variety of attributes. The semi-supervised clustering method requires the user to specify the implementation dependencies between test cases (testing information), which is later combined with coverage (relational information)~\cite{Che11}. The evolving nature of history-based attributes (testing information) has led to the application of reinforcement learning~\cite{Spi17,Wu19}, so the model is able to re-adapt decisions as new samples are provided. Deep learning, which allows building models with memory mechanisms, provides another way to incrementally analyse the past history of test case outcomes~\cite{Xia20}. However, the potential of deep learning to deal with large collections of unstructured data is yet to be explored in the context of TCP. Finally, multi-approaches often combine information attributes from different entities or even groups. The most common approach is to use all the selected attributes to group test cases, then use the same attributes and the class label suggested by the clustering algorithm to feed a classifier that prioritises test cases~\cite{Len13,Rop19}.

\subsection{Differences between industrial and non-industrial ML-based TCP}\label{subsec:rq3_attributes}

Apart from the learning paradigm, differences in the use of information attributes by ML-based TCP methods might respond to the application context, \ie whether the data is collected from industrial or open source systems. Testing information, which was highlighted as the group of attributes most frequently used by industry in Section~\ref{subsec:rq2_attributes}, is also supported by ML works with industrial (8 papers) and non-industrial (8 papers) evaluation (see Table~\ref{tab:attributes_ml}). Nevertheless, it is interesting to note that test case properties (50\%) and similarities (67\%) are often studied from a ML perspective, although they are barely supported by an industrial evaluation (only one paper). In contrast, the suitability of ML techniques to exploit test case history has gained attention from the industry: 67\% of attributes studied in 8 papers, 6 of them with industrial evaluation. Similarly, ML becomes an interesting option to extract knowledge from information available at different moments of the testing process: before executing test cases, dependencies can be automatically analysed (67\% of attributes used); during the execution, time measurements and resource utilisation have become inputs for learning (40\% of attributes within the execution category); and after running test cases, the effectiveness and failure frequency (all attributes from the test case report category) have been used to predict future behaviours of the test cases.

Focusing on SUT-oriented information, almost all code metrics at class level have been considered (92\%), and they are often combined according to the 4 papers found. For instance, estimation of fault-prone classes or code changes has been based on a set of metrics. However, the idea of using ML to detect the parts of the SUT that should be tested first has not been tried in industrial contexts yet. Only one industrial study~\cite{Car11} considers a class metric (\textit{invocations}) with the aim of sorting test cases belonging to the same cluster. Learning from the program inputs has also been explored~\cite{Rop19,Wan12}, but these approaches are validated with open source projects having few parameters of primitive types. How this idea can be extrapolated to industrial projects, which might present more complex parameters, is yet to be investigated.

ML-based TCP methods rely more often on relational information (56\% of ML papers). This is specially evident for coverage and fault-based information if we compare the percentages with those reported in Table~\ref{tab:attributes_ind}. A higher number of coverage formulations have been studied, which are included in a variety of studies: 18 ML studies (52\%) incorporate some kind of coverage function, but only three of them were evaluated in industry (9\%). Change and fault-based attributes, which were highlighted as relevant for industry (67\% and 50\% according to Table~\ref{tab:attributes_ind}), have been studied by three and five ML works, respectively.

\subsection{Considerations when applying machine learning for TCP}\label{subsec:rq3_discussion}

In the context of TCP, instances for ML will mostly correspond to test cases, whereas data features are represented by using information attributes from the \acronym taxonomy as an starting point. In some cases, a direct correspondence is established, \eg the test case execution time becomes a numerical data feature. In other situations, the attribute has to be converted due to the algorithm requirements. As an example, some classifiers cannot deal with categorical values and they require these values to be converted into $n$ binary features (one per category) instead. Most of the available implementations also require that each data feature contains an atomic value, so information attributes of arity 1:N or N:N will be transformed into multiple features too. Other aspects that might influence the final number of features are the number of test cases (\eg for similarity-based attributes), the granularity of the measurement (\eg length of the coverage profile) or the amount of execution history (\eg previous verdicts). Thus, the dimension of the data set prepared for learning can considerably increase.

Table~\ref{tab:features_ml} summarises the minimum, median and maximum number of features found in ML-based TCP methods. If a method has been evaluated with several subsets of features, each combination is considered a different data point in the distribution before computing statistics. Results are grouped by the type of learning approach and the type of evaluation (industrial vs. non-industrial). The number of distinct papers within each category is included too. The last column details whether the number of features varies depending on the SUT under analysis. As an example, all industrially evaluated unsupervised methods have a set of features adapted to the SUT, whereas some of the non-industrially evaluated methods present a fixed number of features, so ``both'' situations can be found. 

\begin{landscape}
\begin{table*}[ht]
\caption{Percentage of attributes considered by TCP methods applying ML.}
\label{tab:attributes_ml}
\scalebox{0.7}{
\begin{tabular}{ll|c|p{1.5cm}p{1.5cm}|p{1.5cm}p{1.5cm}|p{1.5cm}p{1.5cm}|p{1.5cm}p{1.5cm}|p{1.5cm}p{1.5cm}|p{1.5cm}p{1.5cm}}
\toprule
\multirow{2}{*}{\textbf{Group}} & \multirow{2}{*}{\textbf{Entity}} & \multirow{2}{*}{\textbf{\%}} & \multicolumn{2}{c|}{\textbf{Supervised}} & \multicolumn{2}{c|}{\textbf{Unsupervised}} & \multicolumn{2}{c|}{\textbf{Semi-supervised}} & \multicolumn{2}{c|}{\textbf{Reinforcement}} & \multicolumn{2}{c|}{\textbf{Deep learning}} & \multicolumn{2}{c}{\textbf{Multi-approach}}\\
&  &   & Industrial & Non-ind. & Industrial & Non-ind. & Industrial & Non-ind. & Industrial & Non-ind. & Industrial & Non-ind.& Industrial & Non-ind.\\
\toprule
Testing 
& Test case (dependency) & 67\%  & & & \cite{Pra18,Pra19} & &  & \cite{Che11} & & & & & & \\
information 
& Test case (execution) & 40\%  &  & \cite{Ber20}  & \cite{Pra19} &  &  &  & \cite{Spi17} & \cite{Ber20}  &\cite{Xia20}  &  &  & \cite{Bha18}\\
& Test case (history)   & 67\%  & \cite{Bus16} & \cite{Noo17,Pal18} & \cite{Pra18,Pra19} &  &  &  & \cite{Spi17, Wu19} &  &\cite{Xia20}  &  &  & \\
& Test case (property)  & 50\%  & \cite{Bus16} & \cite{Noo17,Pal18} &  & \cite{Kan17} &  &  &  &  & & &  & \\
& Test case (report)    & 100\% & \cite{Bus16} &  & \cite{Ali19} & \cite{Pan13} &  &  & \cite{Spi17, Wu19} &  &\cite{Xia20} & & \cite{Ala18} & \\
& Test case (similarity)& 67\%  &  & \cite{Noo17,Pal18} &  & \cite{Tho14} &  &  &  &  & & &  & \cite{Bha18}\\
& \textit{Total}        & 59\%  &  &  &  &  &  &  &  &  &  & &  & \\
\midrule
SUT 
& Change                & 50\%  &  & \cite{Tan15} &  &  &  &  &  &  &  &  &  &\\
information 
& Class                 & 92\%  & & \cite{Mah20,Sin19,Tou18} & \cite{Car11} &  &  &  &  &  &  &  &  &\\
& Component             & 0\%   &  &  &  &  &  &  &  &  &  &  &  &\\
& Inputs                & 100\% &  & \cite{Wan12} &  & &  &  &  &  &  &  &  & \cite{Rop19}\\
& Program               & 0\%   &  &  &  &  &  &  &  &  &  & & &\\
& \textit{Total}        & 61\%  &  &  &  &  &  &  &  & &  & & &\\
\midrule
Relational 
& Change                & 67\%  & \cite{Bus16} &  & \cite{Ali19} & \cite{Kan17} &  &  &  &  &  & & &\\
information
& Test case (coverage)  & 63\%  & \cite{Lac16} & \cite{Mah20,Noo17,Pal18,Ton06,Wan12} & \cite{Ali19,Car11} & \cite{Che18b,Fan14,Fu17,Leo03,Pan13,Ros17,Yoo09,Zha15} &  & \cite{Che11} &  & \cite{Ngu19} & & &  & \\
& Fault                 & 83\%  & \cite{Lac16} & \cite{Ber20}  & \cite{Car11} & \cite{Fu17} &  &  &  & \cite{Ber20} & & &  & \cite{Len13}\\
& Invocation            & 33\%  &  & \cite{Noo17,Pal18} &  &  &  &  &  & & & &  & \cite{Rop19}\\
& Risk                  & 0\%   &  &  &  &  &  &  &  &  &  & & &\\
& Program               & 75\%  &  & \cite{Jah19} &  &  &  &  &  &  & & & \cite{Ala18} & \cite{Len13,Rop19}\\
& \textit{Total}        & 54\%  &  &  &  &  &  &  &  &  &  & & &\\
\bottomrule
\end{tabular}
}
\end{table*}
\end{landscape}

\begin{table*}[ht]
\caption{Statistics for features used in ML-based TCP studies.}
\label{tab:features_ml}

\begin{tabular}{lcccccc}
\toprule
Type of learning & Industrial eval. & No. Papers & Min. & Median & Max. & SUT-dependent \\
\toprule
Supervised      & Yes & 1*+1 & 5 & 3,502.5 & 3,505 & yes\\
                & No  & 10 & 2 & 8 & 104 & both\\
\midrule
Unsupervised    & Yes & 4 & 4 & 335 & 8,322 & yes\\
                & No  & 10 & 1 & 560 & 36,407 & both\\
\midrule
Semi-supervised & Yes & - & - & - & - & -\\
                & No  & 1 & 236 & N/A & 248 & both\\
\midrule
Reinforcement   & Yes & 2* & 2 & 2.5 & 3 & no\\
                & No  & 2 & 50 & N/A & 50 & both\\
\midrule
Deep learning   & Yes & 1 & 3 & 3 & 3 & no \\
                & No  & - & - & - & - & -\\
\midrule
Multi-approach  & Yes & 1* & 6 & N/A & 6 & no \\
                & No  & 3 & 1 & 3.5 & 5 & both\\
\bottomrule
\end{tabular}
\begin{tablenotes}
\item \footnotesize{Symbol '-' means no method in this category. Symbol '*' refers to papers using industrial datasets only. ``N/A'' stands for not available/not applicable.}
\end{tablenotes}
\end{table*}

As shown in the ``Min'' column, it is possible to find ML studies working with only a small set of features. Such approaches might be more suitable for industry according to the conclusions drawn from Section~\ref{subsec:rq2_combination}. Although the cost of preparing data does not only depend on the number of attributes, it is expected to increase as more features need to be built and maintained. Excluding works using industrial datasets, two papers reporting case studies employ less than 10 features. Those requiring more features correspond to a supervised method that learns from test case descriptions~\cite{Lac16} and two other studies extracting patterns (unsupervised learning) from test case executions. In the former case, a word dictionary is built from the test case specification, each word becoming a feature. Therefore, the final number of features will depend on the word preprocessing and the test suite size. In this context, the aforementioned study presented a test suite of 10,000 test cases that led to 3,500 different words, and consequently more than 3,500 data features. In the latter case, the execution history is divided into testing cycles, each cycle being mapped to one feature. The method was evaluated with systems having between 315 and 8,322 cycles. Here, the objective is to keep only a few rules expressing dependencies between test cases. Reinforcement learning approaches also analyse past executions, but they do not require a long history to start making predictions. The learning policy is dynamically updated after receiving a new fail/pass outcome, so only one feature is modelled and its value updated after each test case execution. A similar concept is applied in the work using DL. As a result, these methods have a low number of features ---between 2 and 3--- and have shown to be effective when learning from industrial datasets.

For a considerable number of works (44\%), the number of features depends on the SUT characteristics or such information is not provided. They are divided into 4 studies with industrial evaluation~\cite{Ali19,Car11,Pra18,Pra19} (3 cases studies and one using a dataset) and 11 studies without industrial evaluation~\cite{Che11,Fan14,Fu17,Kan17,Leo03,Ngu19,Rop19,Ros17,Tho14,Wan12,Zha15}. In the latter case, it would be necessary to study how well it scales if applied to an industrial system with some of the characteristics described in Section~\ref{subsec:rq2_limitations}. Dependency on SUT usually happens for clustering studies based on coverage profiles, as they build a $N$x$M$ similarity matrix, where $N$ is the number of test cases and $M$ is the number of ``code units'' to be covered (data features). Depending on the granularity of such units (statements, methods or blocks), the method will be more or less costly. The presence of a high number of features can result in models that are difficult to comprehend~\cite{Lac16}. Knowing this limitation, some authors evaluate the performance of their algorithms with different subsets of features~\cite{Jah19,Lac16,Len13}. Alternatively, feature selection methods can be executed as part of the data preparation in ML to reduce the dimensionality of the data sets before learning~\cite{Ali19,Che11,Sin19,Ber20}. Both approaches can be highly relevant in industrial contexts to understand the cost-benefit of adding more features, ultimately looking for minimum information requirements, acceptable accuracy and high interpretability. 

Another aspect only affecting supervised techniques is the cost of labelling, \ie assigning a class to each training instance so that its relation with the data features can be established. For TCP, this process differs depending on whether the ML algorithm assigns a priority to each test case or simply predicts its outcome. In case of assigning a priority, previous prioritisation results are needed, what might require manual labelling by experts~\cite{Jah19}. In case of outcome prediction, there is a larger variety of options, including a manual process~\cite{Lac16} and clustering~\cite{Len13}. Semi-supervised techniques have been studied more recently with promising results~\cite{Ala18,Alm16,Rop19}, showing that labelling only a few percentage of samples (10\%) is enough to achieve good classification performance~\cite{Alm16}. It should be noted that the test case verdict (class label) tends to be modelled as a binary variable (fail or pass). Existing industrial datasets makes this simplification, so many ML works do not address more realistic scenarios. Only a few industrial case studies recognise the problem, mentioning that additional ``levels'' in between might be needed~\cite{Ala18} or intentionally omitting issues during test execution~\cite{Bus16,Pra18}. Also, the fact that test case executions are repeated over time implies that the learned model might become obsolete after some time. One single study~\cite{Lac16} poses the need of model update (\ie retraining), with a frequency that would probably depend on project-specific characteristics. Finally, it should be noted that the task of predicting test case effectiveness can suffer from highly imbalanced data, \ie the number of failing test cases is too low in contrast to the positive results. Popular industrial benchmarks, such as the ABB data sets and the Google test suite results, have this problem~\cite{Spi17}. However, very few studies mention this issue and adapt their learning process accordingly~\cite{Ala18, Alm16}.

%--------------------------------
% 8 - DISCUSSION
%--------------------------------
\section{Implications of applying ML for industry-oriented TCP: discussion}\label{sec:discussion}

In this section, we discuss the implications of applying ML for industry-oriented TCP methods. This discussion is presented as a list of topics that arise from the cross-analysis of the information attributes formalised using the \acronym taxonomy (Section~\ref{sec:taxonomy}, focus of RQ1), the analysis of problems encountered in industry (Section~\ref{subsec:rq2_limitations}, related to RQ2) and the observed trends in current ML approaches (Section~\ref{subsec:rq3_discussion}, part of RQ3).

\paragraph{\textbf{Impact of organisational and SUT-specific factors}}
Industrial case studies and our interviews revealed that SUT-specific characteristics and organisational factors are quite important when defining the testing strategy. When these aspects can be captured into information attributes and used for learning, the resulting decision models might be too specific to be reused in other projects. This problem could be mitigated by defining coarse-grained representations for the data features, \eg using categories, instead of the values directly measured from the artefact. Nonetheless, some level of adaptation to the company context seems to be reasonable to address their particular challenges.

\paragraph{\textbf{More representative test case properties}}
A broad variety of values for characteristics of the test cases like its type or status exists in practice. Most of these properties are originally represented as information attributes with binary type, but they could be better represented as categorical variables with a wider range of options. It would provide flexibility to adapt the attribute definition to all the situations that appear in industry: testing levels other than unit testing and changes in the test suite under CI practices, to name two appearing in our interviews with software QA professionals. A broader range of values implies some additional preprocessing (e.g. one-hot encoding) in ML implementations that do not accept categorical features.

\paragraph{\textbf{Role of the testing environment}}
In practice, testing environments present constraints that have proven to influence the design of industry-oriented TCP methods and manual testing strategies according to our literature analysis and interviews with practitioners, respectively. Attributes related to the setup and execution constraints of test cases frequently appear, but are not always considered when applying ML. If time constraints are specially hard, extracting the features and retraining the ML algorithm in each testing cycle might be impracticable. Thinking on how the learning workflow will be integrated into the testing strategy might influence the choice of features and algorithms, as well as their frequency of update.

\paragraph{\textbf{Approximate measurements and proxies}}
Replacing exact measurements, \eg coverage or fault detection capability, by surrogate values or expert's estimates might achieve acceptable performance while alleviating data acquisition effort. This way, the information attributes become static observations easier to handle. Alternatively, reducing the frequency of collection and\slash or analysis can also allow finding a good trade-off between cost and precision. Such a balance is important since having low variety of feature values make it difficult for the ML algorithm to trace boundaries between class labels.

\paragraph{\textbf{Dependency analysis}}
According to the industrial case studies and the conducted interviews, test cases are not always independent units. It could be useful to study the existing dependencies among them or their results, as a way to reduce the number of test cases to be managed. Similar test cases might lead to duplicated instances or correlated features, which will bias the ML-based TCP and unnecessarily increase the overhead if the ML algorithm is costly.

\paragraph{\textbf{Focus on changes}}
If the cost of fully analysing or instrumenting code files is high, extracting information from changes only might be a good alternative. Although change-based information attributes also have an evolving nature and arity N:N, the amount of artefacts is smaller, they can be easily located from commit information, and mapped to what testers use to make decisions. If changes are not tracked or are difficult to collect, then historical results should be considered instead. In case of change-based and history-based attributes, ML paradigms able to continuously digest data (reinforcement learning and deep learning) are more suitable.

\paragraph{\textbf{Learning sensible to test case evolution}}
Changes in the test cases might occur, so the ML approach should be able to adapt to the evolution of the test suite. For attributes analysing similarities via clustering, incremental algorithms~\cite{Bagirov20} could be considered when the arity of the information attribute is affected, \ie test cases are added or removed from the test suite.

\paragraph{\textbf{Non-binary test outcomes}}
Related with the testing process, a binary attribute to represent the test case output might not be representative enough, \eg SUT could crash during testing. Furthermore, the testing management system might contain multiple runs of each test case, with the same or different results. Inconsistent results (\eg flaky tests) could appear too. These circumstances, which were also mentioned during our interviews, should be considered when retrieving data for learning, studying how they affect the definition of data features (which require atomic values) and the class label (binary vs. multiclass).

%--------------------------------
% 9 - THREATS
%--------------------------------
\section{Threats to validity} 
\label{sec:threats}

The bottom-up construction of the \acronym taxonomy and its subsequent analysis pose the following threats to the validity:

\paragraph{Internal validity} Literature search was guided by primary studies selected in secondary studies each one covering a relevant aspect of our work: RT (in particular, TCP), industry relevance and ML application. To avoid missing references not included in these studies but still relevant to our taxonomy, we conducted both manual and automatic searches. The fact that the level of redundancy among sources was low increases the confidence on the adequacy of the final list of references, although it cannot be considered a systematic literature search. In the same line, some relevant TCP papers between 2017 and 2019 could have been excluded due to the selection of certain publication sources. This decision was motivated by the high number of publications retrieved compared to those analysed in the previous years. By focusing on the most renowned sources, we seek to cover the most significant advances in TCP, increasing the chances to find new attributes for inclusion in the taxonomy. Besides, the scope of the analysed publications could be considered limited since we did not included papers addressing test case selection only or grey literature. We focus on TCP since it is more general than test case selection in the sense that prioritising test cases also gives the tester the possibility to select those ranked at the top. Grey literature might provide additional information sources, but the level of detail and validation would be compromised.

\paragraph{Construct validity} The identification and definition of attributes is subject to the authors' analysis of the selected papers, which might contain inaccuracies due to misunderstanding or lack of precise information. To mitigate this, several iterations were performed during the taxonomy development to agree the right level of abstraction for concepts and categories, and multiple values were allowed to cover different approaches. 

\paragraph{External validity} The taxonomy was contrasted against nine recent TCP methods not used for its construction, providing an initial evaluation of its applicability and generalisability. We also presented our work to a short sample of software QA professionals, gathering feedback on its potential use and benefits. From this analysis, the limitations and future evolution of the taxonomy have been discussed.

\paragraph{Conclusion validity} Conclusions about the use of information attributes in industry cannot be considered as fully representative of current practices in industry, since our analysis is restricted to TCP methods published in research literature. The information attributes used might be due to convenience sampling or affected by other types of (selection) biases. To partly mitigate this, we conducted interviews with five software QA professionals from three different companies. They permit us to contrast our analysis with their views, and reinforce the existence of problems associated with TCP automation. Additional interviews to cover a broader variety of companies and TCP practices, and a follow-up of the deployment and application of the taxonomy by testers are future activities that should allow drawing stronger conclusions in this regard.

%--------------------------------
% 10 - CONCLUSION
%--------------------------------
\section{Concluding remarks}
\label{sec:conclusion}

Test case prioritisation plays an important role in RT practices due to large test suites and tight testing cycles. The current spectrum of TCP methods show a variety of techniques and information sources, some of which have been successfully transferred to industry. With the increasing interest in machine learning approaches, which exploit available data from the testing process, the extraction of information relevant for TCP becomes even more important. In this paper, we propose \acronym, a taxonomy to describe the information attributes that have been applied in TCP for the last two decades. We then consider which attributes have been used in industrial studies as well as in studies using ML. The \acronym taxonomy is organised into three dimensions and provides a unified classification to specify not only the source of information, but also other aspects related to the data extraction and measurement process. These dimensions are further divided into seven categories to fully characterise the information attributes, allowing researchers and practitioners to assess their applicability and compare the information needs of their methods.

Based on analysing more than 100 papers, the proposed taxonomy spans 91 information attributes that have been used in TCP studies. We have analysed these attributes from two perspectives in order to understand how they contribute to industrial TCP and ML-oriented methods. We found that studies in industry combines a low number of attributes from a reduced set of sources with a particular focus on past test case outcomes and changes to the SUT. In contrast, the ML-based TCP methods we identified often use a high number of features that can limit their industrial application, specially when they depend on SUT characteristics that might not scale well to industrial systems. Nevertheless, some ML studies analyse and discuss how the data features built from information attributes impact performance. Also, authors are exploring diverse learning approaches to take the historical nature of some attributes into account (\eg reinforcement learning) and overcome lack of information for labelling (\eg semi-supervised learning). We end our study with a discussion of implications to bear in mind when designing ML-based TCP methods suitable for industry. In addition to these type of area-specific analyses, the taxonomy can be used to describe and communicate the information used by new TCP methods. To assist in these tasks, the taxonomy and full details about its use are publicly available from an open repository.

We plan to promote the use of the \acronym taxonomy among researchers and practitioners. We hope that, with the help of the community, periodical revisions of the taxonomy will incorporate new trends in TCP. From our initial interviews with software QA professionals, we have detected that manual TCP is subject to organisational factors and driven by additional information, \eg requirements. Both elements might affect the practical use of the taxonomy since they are difficult to capture as information attributes. Alternatives to include their constraints into automated solutions with ML could be explored in the future as well. Our future research will make use of the taxonomy to guide the experimental study of the cost-effectiveness of data features for industrially applicable ML-based TCP.

\section*{Additional material}\label{sec:additional}

The \acronym taxonomy is hosted in a repository, where the interested reader can navigate through its dimensions, categories and information attributes. Full definitions of all taxonomy elements and supporting references are also available, including the categorisation of those applied by the industry. Finally, the type of learning approach and number of data features are detailed too. The Github repository is available from: \url{http://doi.org/10.5281/zenodo.4400781}.

\begin{acks}

Work supported by the Spanish Ministry of Science and Innovation (project PID2020-115832GB-I00), the University of C\'ordoba (postdoctoral grant ``Plan propio 2019 - mod. 2.4'' and ``Plan propio 2020 - mod. 3.1''), the Andalusian Regional Government (DOC\_00944), and FEDER funds.

\end{acks}

\bibliographystyle{ACM-Reference-Format}
\bibliography{main_tosem}

\end{document}